\documentclass[12pt,nofootinbib,tightenlines]{revtex4}
\usepackage{amsmath,amssymb}
\usepackage{graphicx}
\usepackage{epsfig}
\usepackage{young}
\allowdisplaybreaks

\def \gev          {\ensuremath{\mathrm{\,Ge\kern -0.1em V}}}
\def \mev          {\ensuremath{\mathrm{\,Me\kern -0.1em V}}}

\begin{document}

\title{Lattice QCD study of generalized gluelumps}

\author{Kristen Marsh and Randy Lewis}
\affiliation{Department of Physics and Astronomy, York University, Toronto,
             Ontario, M3J 1P3, Canada}

\begin{abstract}
Proposals for physics beyond the standard model often include new colored
particles at or beyond the scale of electroweak symmetry breaking.  Any
new particle with a sufficient lifetime will bind with standard model gluons
and quarks to form a spectrum of new hadrons.
Here we focus on colored particles in the
octet, decuplet, 27-plet, 28-plet and 35-plet representations of SU(3) color
because these can form hadrons without valence quarks.
In every case, lattice creation operators are constructed for all
angular momentum, parity and charge conjugation quantum numbers.
Computations with fully dynamical lattice QCD configurations produce
numerical results for mass splittings within this new hadron spectrum.
A previous quenched lattice study explored the octet case for certain quantum
number choices, and our findings provide a reassessment
of those early results.
\end{abstract}

\maketitle

\section{Introduction}\label{sec:intro}

Quantum chromodynamics (QCD) describes the interactions between colored
particles such as the color-triplet quarks and color-octet gluons of the
standard model, but additional colored particles are present in many
extensions of the standard model.  Supersymmetry requires gluinos and squarks.
String theory provides a broader range of possibilities.  New strong dynamics
would generate a spectrum of new composite particles (recall the technihadrons of classic
technicolor), and if the new elementary particles (akin to techniquarks) carry QCD color,
then the new composite particles occur as octets, decuplets, and other multiplets of QCD
color.
Studies of new colored particles in the context of the Large Hadron Collider
therefore go far beyond triplets and octets \cite{DelNobile:2009st,delAguila:2010mx,Han:2010rf,Kumar:2011tj,Ilisie:2012cc,Kats:2012ym,Bertuzzo:2012bt},
continuing several decades of interest in the range of color representations
that might be realized beyond the standard model \cite{Ma:1975qy,Karl:1976jk,Wilczek:1976qi,Ng:1978qt,Georgi:1979ng,Dover:1979sn,Marciano:1980zf,Holdom:1982ex,Konishi:1982xh,Lust:1985aw,Braaten:1986eg,Clark:1986vk,Borisov:1986ev,White:1987pq,Fukazawa:1990fb,Chivukula:1990di,Chivukula:1991zk,Kahara:2012yr}.

Lattice QCD is routinely used to obtain quantitative results from the SU(3)
gauge theory of gluons and quarks.  The inclusion of additional particles in an
octet \cite{Kogut:1984sb,Gerstenmayer:1989qw,Karsch:1998qj,Basile:2004wa,Engels:2005te,Cossu:2008wh,Cossu:2009sq},
sextet \cite{Kogut:1984sb,Shamir:2008pb,Kogut:2010cz,Kogut:2011ty,Kogut:2011bd,DeGrand:2012yq,Fodor:2012ty,Sinclair:2012fa,DeGrand:2013uha},
or symmetric \cite{DeGrand:2008kx,DeGrand:2010na}
representation has also been investigated, in some cases applied to a new strong
interaction rather than to QCD itself.
Of more direct relevance to our work is a lattice study by Michael and
coworkers \cite{Campbell:1985kp,Jorysz:1987qj,Michael:1991nc,Foster:1998wu},
culminating in Ref.~\cite{Foster:1998wu} where QCD is coupled to a new heavy
color-octet particle representing the gluino of supersymmetry.
Given that the gluino is significantly heavier than the QCD scale,
Foster and Michael \cite{Foster:1998wu} were able to treat the gluino as
a static particle,
where the spin of the gluino is irrelevant so their results are applicable
more generally to particles of arbitrary spin.
If the static particle is sufficiently stable, then it will couple to
surrounding gluons and quarks to form hadronic bound states.
Foster and Michael used lattice QCD
simulations to produce predictions for mass splittings within this
new spectrum of hadrons.
Specifically, Ref.~\cite{Foster:1998wu} contains numerical results for two
types of hadrons:
gluelumps (having one static octet operator coupled to gluon fields, but no
valence quarks) and adjoint mesons (having one static octet operator coupled to
a quark-antiquark pair).

\begin{table}[tb]
\caption{The smallest gluelump mass splittings relative to the $1^{+-}$ state
from the original lattice simulation \cite{Foster:1998wu} (where errors are
statistical only), compared to model calculations published subsequently.
See Sec.~\ref{sec:results} for a crucial discussion of lattice systematics.
See the original publications for detailed discussions about other parameter
choices and systematic issues; this table is merely an introduction.
(To display data from Ref.~\cite{Buisseret:2008pd} we chose $r_0=0.5$ fm.)}
\label{table:FMresults}
\begin{tabular}{llcccc} 
\hline
\hline
$J^{PC}$~~~ & \multicolumn{5}{c}{$M(J^{PC})-M(1^{+-})$~~ [GeV]} \\
\cline{2-6}
         & Lattice \cite{Foster:1998wu} & ~~~Bag \cite{Karl:1999wq}
         & ~~~String \cite{Simonov:2000ky} & ~~~Coulomb gauge \cite{Guo:2007sm}
         & ~~~Transverse gluons \cite{Buisseret:2008pd} \\
\hline
$1^{--}$ & 0.368(7)  & 0.55       & 0.47 & 0.40 & 0.37 \\
$2^{--}$ & 0.567(10)\footnote{This entry repairs a simple typo in column 4 of
Table III in Ref.~\cite{Foster:1998wu}, as can be seen by comparing with
column 3 of that same table and with Fig.~3 in Ref.~\cite{Foster:1998wu}.}
                     & 0.54       & 0.49 & 0.59 & 0.57 \\
$3^{+-}$ & 0.972(24) & 1.01       & 0.84 & 1.11 & 0.97 \\
$2^{+-}$ & 0.973(36) & 1.21       & 0.83 & 0.71 & 0.94 \\
$0^{++}$ & 1.092(28) & $\sim$1.2  & 0.91 & $\cdots$ & $\cdots$ \\
\hline
\hline
\end{tabular}
\end{table}
According to Ref.~\cite{Foster:1998wu}, the lightest gluelump has
$J^{PC}=1^{+-}$.  The predicted
mass splittings of the five next-lightest gluelumps are shown
in Table~\ref{table:FMresults}.  Four model calculations
\cite{Karl:1999wq,Simonov:2000ky,Guo:2007sm,Buisseret:2008pd} are also shown in
Table~\ref{table:FMresults} for comparison.
We display mass differences because these are what emerge directly from the
lattice simulations, but in fact
the absolute mass scale has been determined in Ref.~\cite{Bali:2003jq}
using a combination of effective field theory and related lattice QCD input.
After fixing this absolute mass scale, Ref.~\cite{Bali:2003jq} then takes
the gluelump mass splittings directly from Ref.~\cite{Foster:1998wu}.
We point to potential NRQCD \cite{Brambilla:1999xf} as an example of an
important theoretical development that has requested further lattice studies
of gluelumps.

The authors of Ref.~\cite{Foster:1998wu} expressed surprise at the heaviness
of their $0^{++}$ state, and also at the degeneracy of $2^{+-}$ and $3^{+-}$.
Lattice simulations use irreducible representations $\Lambda$
of the octahedral group
rather than continuum angular momentum $J$, so, for example, a $J=2$ state should
appear for both $\Lambda=E$ and $\Lambda=T_2$,
but Ref.~\cite{Foster:1998wu} points
out that $E^{++}$ and $T_2^{++}$ are not degenerate in their lattice
data though the discrepancy is consistent with degeneracy in the continuum
limit.  Because of the computational expense, Ref.~\cite{Foster:1998wu} made use
of quenched lattices so the authors expect at least a 10\% systematic error.
The work also relied exclusively on operators built from square paths on the
lattice which allows access to only half of the possible $\Lambda^{PC}$
representations (i.e.\ 10 out of 20), leaving quantum numbers such as
$J^{PC}=0^{+-}$, $0^{-+}$, $0^{--}$, and $1^{++}$ unstudied.

In the present work, we extend the basis of operators to the complete set of
$\Lambda^{PC}$ options, and we use dynamical (unquenched) lattices.
This provides an opportunity to revisit some of the surprises revealed
by Foster and Michael in their seminal work, and to predict additional
gluelump masses.
We also develop operators for generalized gluelumps by replacing
the static octet source with a static source having a larger color
representation.  To avoid the expense of lattice simulations with valence
quarks, we choose representations that need only gluons to produce a
color-singlet generalized gluelump.  Specifically we choose dimensions 10, 27,
28, and 35.  We reiterate that our numerical results make use of dynamical
lattice simulations so that virtual quarks and antiquarks are retained.

Static propagators are known to produce particularly large statistical
uncertainties in lattice simulations, and a static octet particle is noisier
than a static triplet \cite{Michael:1985ne,Foster:1998wu}.
We expect that the larger representations included in the present study will be
noisier still.
Also, the Casimir scaling hypothesis \cite{Ambjorn:1984mb,DelDebbio:1995gc,Bali:2000gf}
is the notion that the string tension between strongly interacting particles
should be proportional to the quadratic Casimir, and
standard group theory \cite{Lichtenberg:1991,Georgi:1982jb} shows that the
quadratic Casimirs for our representations,
normalized such that the triplet has $C_2(3)=4/3$, are $C_2(8)=3$,
$C_2(10)=6$, $C_2(27)=8$, $C_2(28)=18$, and $C_2(35)=12$.
Polyakov loops with all of these representations have been tested previously for Casimir scaling:
see Table 2 of Ref.~\cite{Mykkanen:2012ri}.  [For other lattice studies of
Casimir scaling and various representations in four-dimensional SU(3) gauge theory,
sometimes in the context of $n$-ality, see
Refs.~\cite{Deldar:1999vi,Bali:2000un,Lucini:2001ej,DelDebbio:2002yp,Dumitru:2003hp,DelDebbio:2003tk,Gupta:2007ax,Anzai:2010td,Greensite:2011gg,Lucini:2012gg}.
The present work deals exclusively with zero $n$-ality.]
In the case of gluelumps, our simulations confirm that signals for representations with larger Casimirs
are damped more
rapidly as a function of Euclidean time, as well as being statistically noisy.
Despite these substantial difficulties, the numerical results of this project
provide useful information about representations beyond the octet,
as well as the octet itself.

\section{Correlation functions}\label{sec:correlators}

Generalized gluelumps do not involve valence quarks, so the heavy static
particle must be able to form a color singlet by coupling to a collection of
octet gauge fields,
\begin{equation}
8 \otimes 8 \otimes 8 \otimes \cdots ~~~\in~~~
1 \oplus 8 \oplus 10 \oplus \overline{10} \oplus 27 \oplus 28 \oplus
\overline{28} \oplus 35 \oplus \overline{35} \oplus \cdots \,.
\end{equation}
Representations of dimension $n_D=8$, 10, 27, 28, and 35 will be considered in
this work.  The corresponding Young tableaux, derivable using standard group
theory methods \cite{Lichtenberg:1991,Georgi:1982jb}, are displayed in
Table~\ref{table:young}.
\begin{table}[tb]
\caption{Young tableaux for representations relevant to this work.  Labels
inside boxes are to aid the discussion of (anti)symmetrization.}
\label{table:young}
\begin{tabular}{c|ccccc}
\hline\hline
$n_D$ & 8 & 10 & 27 & 28 & 35 \\
\hline \vspace{-5mm} \\
Tableau & \begin{Young} i&j\cr k\cr \end{Young}
        & \begin{Young} i&j&k\cr \end{Young}
        & \begin{Young} i&j&k&l\cr m&n\cr \end{Young}
        & \begin{Young} i&j&k&l&m&n\cr \end{Young}
        & \begin{Young} i&j&k&l&m\cr n\cr \end{Young} \\
\hline\hline
\end{tabular}
\end{table}
Notice that the number of boxes in each tableau is a multiple of 3, as
required for tableaux built exclusively from octet gauge fields.
As will be discussed below, each generalized gluelump will have a tensor
where the number of indices equals the number of columns in its Young tableau.

As is standard in lattice QCD simulations, mass splittings will be obtained
by computing a correlation function and then
observing the exponential dependence on Euclidean time.
A correlation function that creates a gluelump at Euclidean time $\tau_i$ and
then annihilates it at time $\tau_f$ is
\begin{equation}\label{correlator}
C(\tau_f-\tau_i) = H^{(n_D) \alpha \dagger} (\tau_i)
                   G^{(n_D) \alpha \beta} (\tau_i,\tau_f)
                   H^{(n_D) \beta} (\tau_f) \, .
\end{equation}
Repeated indices $\alpha$ and $\beta$ are summed from 1 to $n_D$ to produce a
gauge-invariant correlation function.
The operators $H$ and $H^\dagger$ that, respectively, annihilate and create the
required gauge field structure will be developed in Sec.~\ref{sec:operators}.
The propagator $G$ for the static particle is described presently.

\subsection{Static propagator}\label{sec:propagators}

A static particle propagates purely in the temporal direction (subscript ``4''),
so for a representation of dimension $n_D$ we can write
\begin{equation}
G^{(n_D) \alpha \beta}(\tau_i,\tau_f)
 = U^{(n_D) \alpha \gamma}_{4}(\vec{x},\tau_i)
   U^{(n_D) \gamma \delta}_{4}(\vec{x},\tau_i+a)
   U^{(n_D) \delta \epsilon}_{4}(\vec{x},\tau_i+2a) \cdots
   U^{(n_D) \zeta \beta}_{4}(\vec{x},\tau_f)
\end{equation}
with repeated Greek indices summed from 1 to $n_D$.
Each generalized link $U^{(n_D)}$ is built from elementary links (one per
column of the Young tableau) contracted at each end (e.g.\ Euclidean times
$\tau_i$ and $\tau_i+a$) with a basis tensor $T$,
\begin{eqnarray}
U^{(8) \alpha \beta}
   &=& U_{ik} U^*_{jl} T_{ij}^{\alpha} T_{kl}^{\beta} \,, \\
U^{(10) \alpha \beta}
   &=& U_{il} U_{jm} U_{kn} T_{ijk}^{\alpha} T_{lmn}^{\beta} \,, \\
U^{(27) \alpha \beta}
   &=& U_{im} U_{jn} U^*_{ko} U^*_{lp} T_{ijkl}^\alpha T_{mnop}^{\beta} \,, \\
U^{(28) \alpha \beta}
   &=& U_{io} U_{jp} U_{kq} U_{lr} U_{ms} U_{nt} T_{ijklmn}^\alpha T_{opqrst}^{\beta} \,, \\
U^{(35) \alpha \beta}
   &=& U_{in} U_{jo} U_{kp} U_{lq} U^*_{mr} T_{ijklm}^\alpha T_{nopqr}^{\beta} \,,
\end{eqnarray}
where repeated color indices $i,j,k,\ldots$ are summed from 1 to 3.

An acceptable basis for the octet representation is
$T^\alpha=\lambda^\alpha/\sqrt{2}$ where $\lambda^\alpha$ is a standard
Gell-Mann matrix as was used in Ref.~\cite{Foster:1998wu}.  Beyond the
octet we find it more convenient to use real $T$ tensors, and for
consistency we will also use real matrices for the octet itself.

For a Young tableau with $n_B$ boxes, we begin with an arbitrary tensor
having $n_B$ indices.  Then we symmetrize all indices within a row.  Next
we antisymmetrize all indices within any column having two boxes and multiply
that pair of indices by a Levi-Civit\`{a} tensor, thus reducing the number of
indices by one for each antisymmetrized column.
The final step is to select a real basis of $T$ tensors.
For example, consider the 27-plet.  An arbitrary 6-index tensor is
$a^{ijklmn}$, and after symmetrization of $(i,j,k,l)$, symmetrization of $(m,n)$
and then antisymmetrization of $(i,m)$ and $(j,n)$, we have
\begin{equation}
b^{ijklmn} = a^{ijklmn} - a^{mjklin} - a^{inklmj} + a^{mnklij} + \cdots
\end{equation}
which reduces to a 4-index tensor,
\begin{equation}
T_{klpq} = \frac{1}{4}\epsilon_{imp}\epsilon_{jnq}b^{ijklmn} \,.
\end{equation}
Evaluation of all $3^4=81$ elements of this tensor reveals that it contains
36 distinct entries but only 27 of them are linearly independent due to the
following 9 constraints:
\begin{eqnarray}
T_{1112}+T_{1222}+T_{1323}=0 \,, \nonumber \\
T_{1211}+T_{2212}+T_{2313}=0 \,, \nonumber \\
T_{1111}+T_{1212}+T_{1313}=0 \,, \nonumber \\
T_{1113}+T_{1333}+T_{1223}=0 \,, \nonumber \\
T_{2322}+T_{3323}+T_{1312}=0 \,, \nonumber \\
T_{2222}+T_{1212}+T_{2323}=0 \,, \nonumber \\
T_{2223}+T_{2333}+T_{1213}=0 \,, \nonumber \\
T_{3313}+T_{1311}+T_{2312}=0 \,, \nonumber \\
T_{3333}+T_{1313}+T_{2323}=0 \,.
\end{eqnarray}
Our choice for the basis of 27 tensors is given explicitly in
Appendix~\ref{app:tensors}
together with the other representations: octet, decuplet, 28-plet, and 35-plet.

As a useful check of these expressions, we calculate a completeness relation
for each case: Appendix~\ref{app:tensors} verifies that the quantity
\begin{equation}
\sum_{\alpha=1}^{n_D}T^\alpha T^\alpha
\end{equation}
comprises a simple Kronecker delta structure.  This is important for the
gauge invariance of our correlation functions.

Notice also that our decuplet representation agrees with Appendix B of
Ref.~\cite{DelDebbio:2008zf}.
Finally, we mention that we have verified numerically that our real basis of
octet $T$ tensors produces correlation functions that are identical to
those obtained in the Gell-Mann basis.

\subsection{Creation/annihilation operators}\label{sec:operators}

The remaining ingredient needed for the computation of correlation functions
is the set of operators, $H$ of Eq.~(\ref{correlator}), coupling to the
generalized gluelumps.  An $H$ operator is built from products of gauge links
that join to the static particle propagator via a $T$ tensor
(from Sec.~\ref{sec:propagators}).
Planar squares were used for $H$ operators in Ref.~\cite{Foster:1998wu}, but
this provides access to only half of the possible quantum numbers.  Our most basic
building block will be a ``chair,'' i.e.\ a 1$\times$2 rectangle bent to a
90$^\circ$ angle, which provides access to all quantum numbers.
For extra confirmation of numerics, we also ran simulations with the planar
square operators used by Foster and Michael, and we verified that results are
consistent with the corresponding chair-based operators defined here.

\begin{figure}[htp]
\includegraphics[width=16cm,clip=true]{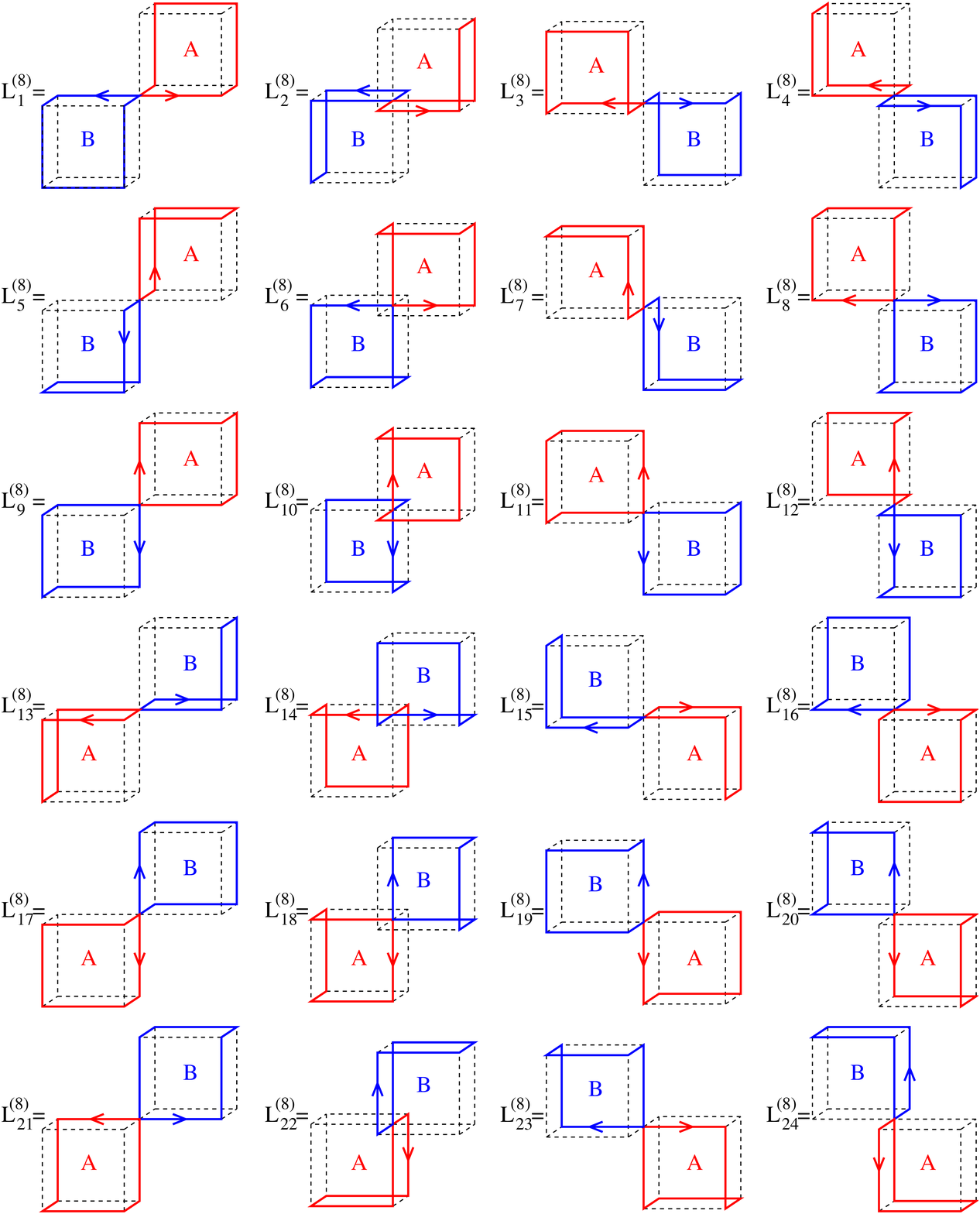}
\caption{Each chair-shaped path is the product of six gauge links used to build
operators for
octet gluelumps.  Solid lines are the gauge links; dashed lines are just to
aid with three-dimensional (3D) visualization.}
\label{fig:chairs8} 
\end{figure}
Figure~\ref{fig:chairs8} displays a pair of chairs touching each other
at one lattice site
and rotated into all of the 24 orientations that are possible on a cubic
lattice.  Notice that each chair has a particular direction because a 
``backward link'' $U_{-\mu}(x+\mu)=U^\dagger_\mu(x)$ is not equal to the
``forward link'' $U_\mu(x)$.
Within each pair of chairs in Fig.~\ref{fig:chairs8}, $A+B$ is a positive
parity operator and
$A-B$ is a negative parity operator.  Because $U_\mu(x)\to U^\dagger_\mu(x)$
under charge conjugation, a ``forward'' chair plus a ``backwards'' chair has positive charge
conjugation and the difference between these two chairs has negative charge conjugation.

The five bosonic irreducible representations of the octahedral group are
$\Lambda=A_1$, $A_2$,
$T_1$, $T_2$, and $E$, and their smallest continuum angular momenta are
$J=0$, 3, 1, 2, and 2, respectively.
For octet gluelumps, the corresponding operators are obtained
from specific linear combinations of
the chair-shaped paths in Fig.~\ref{fig:chairs8}.  The steps of a
derivation are provided in Appendix~\ref{app:operators},
and the results are given here:
\begin{eqnarray}
H^{(8) \alpha}(A_1) &=& \left( \sum_{a=1}^{24} L_a^{(8)} \right)_{ij} T_{ij}^\alpha \,, \nonumber \\
H^{(8) \alpha}(A_2) &=& \left( \sum_{a=1}^{12} (-1)^a L_a^{(8)} - \sum_{a=13}^{24} (-1)^a L_a^{(8)} \right)_{ij} T_{ij}^\alpha \,, \nonumber \\
H^{(8) \alpha}(T_1^x) &=& \left( L_{6}^{(8)} + L_{20}^{(8)} + L_{21}^{(8)} + L_{11}^{(8)} - L_{18}^{(8)} - L_{8}^{(8)} - L_{9}^{(8)} - L_{23}^{(8)} \right)_{ij} T_{ij}^\alpha \,, \nonumber \\
H^{(8) \alpha}(T_1^y) &=& \left( L_{5}^{(8)} + L_{19}^{(8)} + L_{24}^{(8)} + L_{10}^{(8)} - L_{17}^{(8)} - L_{7}^{(8)} - L_{12}^{(8)} - L_{22}^{(8)} \right)_{ij} T_{ij}^\alpha \,, \nonumber \\
H^{(8) \alpha}(T_1^z) &=& \left( L_{1}^{(8)} + L_{2}^{(8)} + L_{3}^{(8)} + L_{4}^{(8)} - L_{13}^{(8)} - L_{14}^{(8)} - L_{15}^{(8)} - L_{16}^{(8)} \right)_{ij} T_{ij}^\alpha \,, \nonumber \\
H^{(8) \alpha}(T_2^x) &=& \left( L_{6}^{(8)} - L_{20}^{(8)} + L_{21}^{(8)} - L_{11}^{(8)} + L_{18}^{(8)} - L_{8}^{(8)} + L_{9}^{(8)} - L_{23}^{(8)} \right)_{ij} T_{ij}^\alpha \,, \nonumber \\
H^{(8) \alpha}(T_2^y) &=& \left( L_{5}^{(8)} - L_{19}^{(8)} + L_{24}^{(8)} - L_{10}^{(8)} + L_{17}^{(8)} - L_{7}^{(8)} + L_{12}^{(8)} - L_{22}^{(8)} \right)_{ij} T_{ij}^\alpha \,, \nonumber \\
H^{(8) \alpha}(T_2^z) &=& \left( L_{1}^{(8)} - L_{2}^{(8)} + L_{3}^{(8)} - L_{4}^{(8)} + L_{13}^{(8)} - L_{14}^{(8)} + L_{15}^{(8)} - L_{16}^{(8)} \right)_{ij} T_{ij}^\alpha \,, \nonumber \\
H^{(8) \alpha}(E^1) &=& \left( v^x - v^y \right)_{ij} T_{ij}^\alpha \,, \nonumber \\
H^{(8) \alpha}(E^2) &=& \left( v^x + v^y - 2 v^z \right)_{ij} T_{ij}^\alpha \,, \nonumber \\
v^x &=&  L_{6}^{(8)} + L_{20}^{(8)} + L_{21}^{(8)} + L_{11}^{(8)} + L_{18}^{(8)} + L_{8}^{(8)} + L_{9}^{(8)} + L_{23}^{(8)}  \,, \nonumber \\
v^y &=&  L_{5}^{(8)} + L_{19}^{(8)} + L_{24}^{(8)} + L_{10}^{(8)} + L_{17}^{(8)} + L_{7}^{(8)} + L_{12}^{(8)} + L_{22}^{(8)}  \,, \nonumber \\
v^z &=&  L_{1}^{(8)} + L_{2}^{(8)} + L_{3}^{(8)} + L_{4}^{(8)} + L_{13}^{(8)} + L_{14}^{(8)} + L_{15}^{(8)} + L_{16}^{(8)}  \,.
\label{eq:operators}
\end{eqnarray}
Notice that $A_1$ and $A_2$ are one-dimensional representations,
$T_1$ and $T_2$ are three dimensional, and $E$ is two dimensional.

\begin{figure}[tb]
\includegraphics[width=10cm,clip=true]{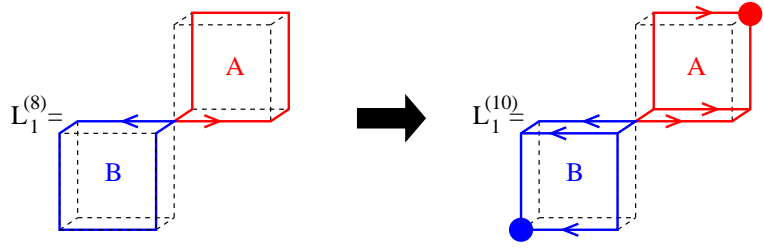}
\caption{Octet chairs and decuplet chairs have the same shape but the product
of gauge links differs.
Solid lines are the gauge links; dashed lines are just to
aid with 3D visualization.
A filled circle denotes insertion of a Levi-Civit\`{a} tensor.}
\label{fig:chairs10} 
\end{figure}
Each decuplet chair contains three paths that begin at a central lattice site
(where the tensor $T$ will be placed) and end at a Levi-Civit\`{a} tensor.
Each of those three paths is the product of three gauge links.
The precise definition of $L_1^{(10)}$ is displayed in Fig.~\ref{fig:chairs10},
and $L_2^{(10)}$ through $L_{24}^{(10)}$ are defined by applying the same
procedure to every chair in Fig.~\ref{fig:chairs8}.
The decuplet operators are obtained by making two simple adjustments to
Eqs.~(\ref{eq:operators}): replace every superscript $(8)$ with a
superscript $(10)$
and replace every pair of indices $ij$ by the three indices $ijk$.

The 35-plet is built from a double chair, specifically one octet-type chair
and one decuplet-type chair, defined as follows:
\begin{eqnarray}
\left( L_{ 1}^{(35)} \right)_{ijklm} = \left( L_{ 5}^{(8)} \right)_{im} \left( L_{ 9}^{(10)} \right)_{jkl} \,, & \qquad &
\left( L_{13}^{(35)} \right)_{ijklm} = \left( L_{21}^{(8)} \right)_{im} \left( L_{17}^{(10)} \right)_{jkl} \,, \nonumber \\
\left( L_{ 2}^{(35)} \right)_{ijklm} = \left( L_{ 6}^{(8)} \right)_{im} \left( L_{10}^{(10)} \right)_{jkl} \,, & \qquad &
\left( L_{14}^{(35)} \right)_{ijklm} = \left( L_{22}^{(8)} \right)_{im} \left( L_{18}^{(10)} \right)_{jkl} \,, \nonumber \\
\left( L_{ 3}^{(35)} \right)_{ijklm} = \left( L_{ 7}^{(8)} \right)_{im} \left( L_{11}^{(10)} \right)_{jkl} \,, & \qquad &
\left( L_{15}^{(35)} \right)_{ijklm} = \left( L_{23}^{(8)} \right)_{im} \left( L_{19}^{(10)} \right)_{jkl} \,, \nonumber \\
\left( L_{ 4}^{(35)} \right)_{ijklm} = \left( L_{ 8}^{(8)} \right)_{im} \left( L_{12}^{(10)} \right)_{jkl} \,, & \qquad &
\left( L_{16}^{(35)} \right)_{ijklm} = \left( L_{24}^{(8)} \right)_{im} \left( L_{20}^{(10)} \right)_{jkl} \,, \nonumber \\
\left( L_{ 5}^{(35)} \right)_{ijklm} = \left( L_{ 9}^{(8)} \right)_{im} \left( L_{ 1}^{(10)} \right)_{jkl} \,, & \qquad &
\left( L_{17}^{(35)} \right)_{ijklm} = \left( L_{13}^{(8)} \right)_{im} \left( L_{21}^{(10)} \right)_{jkl} \,, \nonumber \\
\left( L_{ 6}^{(35)} \right)_{ijklm} = \left( L_{10}^{(8)} \right)_{im} \left( L_{ 2}^{(10)} \right)_{jkl} \,, & \qquad &
\left( L_{18}^{(35)} \right)_{ijklm} = \left( L_{14}^{(8)} \right)_{im} \left( L_{22}^{(10)} \right)_{jkl} \,, \nonumber \\
\left( L_{ 7}^{(35)} \right)_{ijklm} = \left( L_{11}^{(8)} \right)_{im} \left( L_{ 3}^{(10)} \right)_{jkl} \,, & \qquad &
\left( L_{19}^{(35)} \right)_{ijklm} = \left( L_{15}^{(8)} \right)_{im} \left( L_{23}^{(10)} \right)_{jkl} \,, \nonumber \\
\left( L_{ 8}^{(35)} \right)_{ijklm} = \left( L_{12}^{(8)} \right)_{im} \left( L_{ 4}^{(10)} \right)_{jkl} \,, & \qquad &
\left( L_{20}^{(35)} \right)_{ijklm} = \left( L_{16}^{(8)} \right)_{im} \left( L_{24}^{(10)} \right)_{jkl} \,, \nonumber \\
\left( L_{ 9}^{(35)} \right)_{ijklm} = \left( L_{ 1}^{(8)} \right)_{im} \left( L_{ 5}^{(10)} \right)_{jkl} \,, & \qquad &
\left( L_{21}^{(35)} \right)_{ijklm} = \left( L_{17}^{(8)} \right)_{im} \left( L_{13}^{(10)} \right)_{jkl} \,, \nonumber \\
\left( L_{10}^{(35)} \right)_{ijklm} = \left( L_{ 2}^{(8)} \right)_{im} \left( L_{ 6}^{(10)} \right)_{jkl} \,, & \qquad &
\left( L_{22}^{(35)} \right)_{ijklm} = \left( L_{18}^{(8)} \right)_{im} \left( L_{14}^{(10)} \right)_{jkl} \,, \nonumber \\
\left( L_{11}^{(35)} \right)_{ijklm} = \left( L_{ 3}^{(8)} \right)_{im} \left( L_{ 7}^{(10)} \right)_{jkl} \,, & \qquad &
\left( L_{23}^{(35)} \right)_{ijklm} = \left( L_{19}^{(8)} \right)_{im} \left( L_{15}^{(10)} \right)_{jkl} \,, \nonumber \\
\left( L_{12}^{(35)} \right)_{ijklm} = \left( L_{ 4}^{(8)} \right)_{im} \left( L_{ 8}^{(10)} \right)_{jkl} \,, & \qquad &
\left( L_{24}^{(35)} \right)_{ijklm} = \left( L_{20}^{(8)} \right)_{im} \left( L_{16}^{(10)} \right)_{jkl} \,.
\label{eq:chairs35}
\end{eqnarray}
Notice that a diagram of $L_n^{(35)}$ would resemble $L_n^{(8)}$ of
Fig.~\ref{fig:chairs8} except that the chairs in each lattice cell are in
the opposite locations (there {\em is} a chair where there {\em was not},
and there {\em is not} a chair where there {\em was}).
There are two options---the octet-type chair could have been to the left or
to the right of the decuplet-type chair when viewed from a certain angle---and 
Eqs.~(\ref{eq:chairs35}) show which of the two options we have selected.
The 35-plet operators are obtained by making two simple adjustments to
Eqs.~(\ref{eq:operators}): replace every superscript $(8)$ with a
superscript $(35)$
and replace every pair of indices $ij$ by the set $ijklm$.

The 27-plet is also built from a double chair, but both are octet-type chairs.
The definition is obtained from Eqs.~(\ref{eq:chairs35}) with these
replacements: $(35)\to(27)$, $(10)\to(8)$, $ijklm\to ijkl$, $im\to ik$,
$jkl\to jl$.
The 27-plet operators are obtained by making two simple adjustments to
Eqs.~(\ref{eq:operators}): replace every superscript $(8)$ with a
superscript $(27)$
and replace every pair of indices $ij$ by the set $ijkl$.

The 28-plet is built from a double chair; both are decuplet-type.
The definition is obtained from Eqs.~(\ref{eq:chairs35}) with these
replacements: $(35)\to(28)$, $(8)\to(10)$, $ijklm\to ijklmn$, $im\to ijk$,
$jkl\to lmn$.
The 28-plet operators are obtained by making two simple adjustments to
Eqs.~(\ref{eq:operators}): replace every superscript $(8)$ with a
superscript $(28)$
and replace every pair of indices $ij$ by the set $ijklmn$.

To complete the discussion of generalized gluelump operators, notice that
the octet and 27-plet are eigenstates of charge conjugation, whereas the
decuplet, 28-plet, and 35-plet are not.  This is evident from the Young tableaux representations of the underlying group theory as shown in Table \ref{table:young}.  Representations with twice as many boxes in the top row as the bottom row have the same number of symmetric and antisymmetric indices.  They are their own antirepresentations and are eigenstates of charge conjugation.  Other representations are ``charged'' and cannot form states with definite charge conjugation.  This property can also be seen in the color flow
in Figs.~\ref{fig:chairs8} and \ref{fig:chairs10}.  Any single octet chair
has one color and one anticolor emanating from the central lattice site,
but the decuplet has three colors and no anticolors.

\section{Lattice simulations}\label{sec:results}

The simulations performed for this work use two ensembles of configurations
provided by the CP-PACS and JLQCD Collaborations \cite{Ishikawa:2007nn}.
These ensembles are $O(a)$-improved due to the use of the clover coefficient,
$c_{\rm SW}$.  The lattice spacings are comparable to the smallest values used
in Ref.~\cite{Foster:1998wu}.  Precise parameter values are displayed in
Table~\ref{table:configs}.
Notice that the strange quark mass is essentially its physical value, but
the up and down quarks are not: the pion is about 3.5 times heavier than its
physical value.
\begin{table}[tb]
\caption{Input parameters and standard output parameters (separated by a
horizontal line) used in this work were obtained from
Ref.~\cite{Ishikawa:2007nn}.
For comparison, parameters used in the quenched study by
Ref.~\cite{Foster:1998wu} are also shown.}\label{table:configs}
\begin{tabular}{ccccccc}
\hline
\hline
Source          & \cite{Ishikawa:2007nn} & \cite{Ishikawa:2007nn}
                & \cite{Foster:1998wu} & \cite{Foster:1998wu}
                & \cite{Foster:1998wu} \\
$\beta$         & 1.90 & 2.05 & 5.7 & 6.0 & 6.2 \\
$\kappa_{ud}$   & 0.13700 & 0.13560 & $\cdots$ & $\cdots$ & $\cdots$ \\
$\kappa_s$      & 0.13640 & 0.13540 & $\cdots$ & $\cdots$ & $\cdots$ \\
$c_{\rm SW}$    & 1.7150 & 1.6280 & $\cdots$ & $\cdots$ & $\cdots$ \\
$L^3\times T$   & $20^3\times40$ & $28^3\times56$ & $12^3\times24$
                & $16^3\times48$ & $24^3\times48$ \\
No. of configurations  & 790 & 650 & 99 & 202 & 60 \\
\hline
Lattice spacing [fm] & 0.0982(19) & 0.0685(26) & 0.170 & 0.0948 & 0.0683 \\
$m_\pi/m_\rho$  & 0.6243(28) & 0.6361(47) & $\cdots$ & $\cdots$ & $\cdots$ \\
$\sqrt{2m_K^2-m_\pi^2}/m_\phi$  & 0.7102(20) & 0.6852(46) & $\cdots$ & $\cdots$ & $\cdots$ \\
\hline
\hline
\end{tabular}
\end{table}

\begin{figure}[tb]
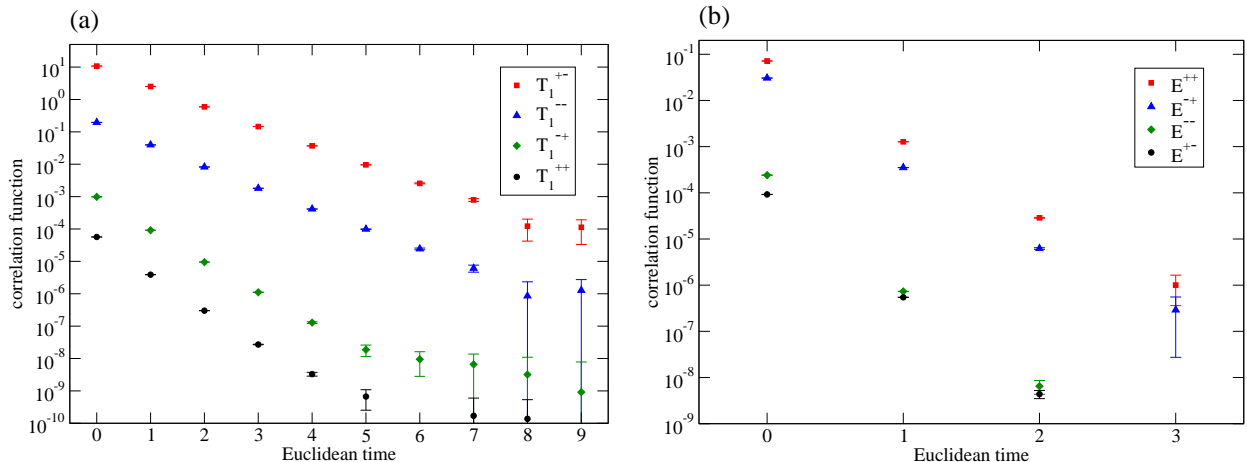

\includegraphics[angle=0,width=8cm,trim=0 0 0 0,clip=true]{octet205.eps}
\hfill
\includegraphics[angle=0,width=8cm,trim=0 0 0 0,clip=true]{27plet190.eps}
\caption{Sample correlation functions:
(a) the $T_1^{PC}$ channels containing a static octet particle at $\beta=2.05$,
(b) the $E^{PC}$ channels containing a static 27-plet particle at $\beta=1.90$.}
\label{fig:8correlators} 
\end{figure}

Stout link smearing \cite{Morningstar:2003gk} was applied to the operators of
Sec.~\ref{sec:operators} with parameters tuned to reduce contamination from
excited states.  In the notation of Ref.~\cite{Morningstar:2003gk}, we use
$(\rho,n_\rho)=(0.20,15)$ for the octet and decuplet, and we use
$(\rho,n_\rho)=(0.15,15)$ for the 27-plet, 28-plet, and 35-plet.
Figure~\ref{fig:8correlators} gives an indication of the quality of the data
by showing the relatively clean example of the octet $T_1^{PC}$ as well as
the much more challenging example of the 27-plet $E^{PC}$.

Mass differences are obtained by the simultaneous fit of a pair of correlation
functions:
\begin{equation}
C_1 = f_1e^{-M_1\tau}
~~~~~{\rm and}~~~~~
C_2 = f_2e^{-(M_1+\delta M_{12})\tau}
\end{equation}
where $f_1$, $f_2$, $M_1$, and $\delta M_{12}$ are the four fit parameters.
The mass difference $\delta M_{12}$ is the physics we wish to extract, and its
statistical uncertainty is determined by bootstrapping \cite{Efron:1993}.
The most important systematic uncertainty comes from choosing the range of
time steps, $\tau_i$ to $\tau_f$, to include in each fit. 
Fits do not depend significantly on $\tau_f$ because the inclusion of noisy
data at large Euclidean times has a negligible influence.
We determined the range of $\tau_i$ options that all produced a common
$\delta M_{12}$ value within one statistical standard deviation, and then used
the smallest $\tau_i$ in that range because it produces the smallest statistical
uncertainty.  A one-sigma systematic error was then assigned to be
\begin{equation}\label{systematicerror}
\left|\frac{\delta M_{12}(\tau_i)-\delta M_{12}(\tau_i-1)}{2}\right| \,.
\end{equation}

\begin{figure}[tb]
\includegraphics[clip=true,height=10cm]{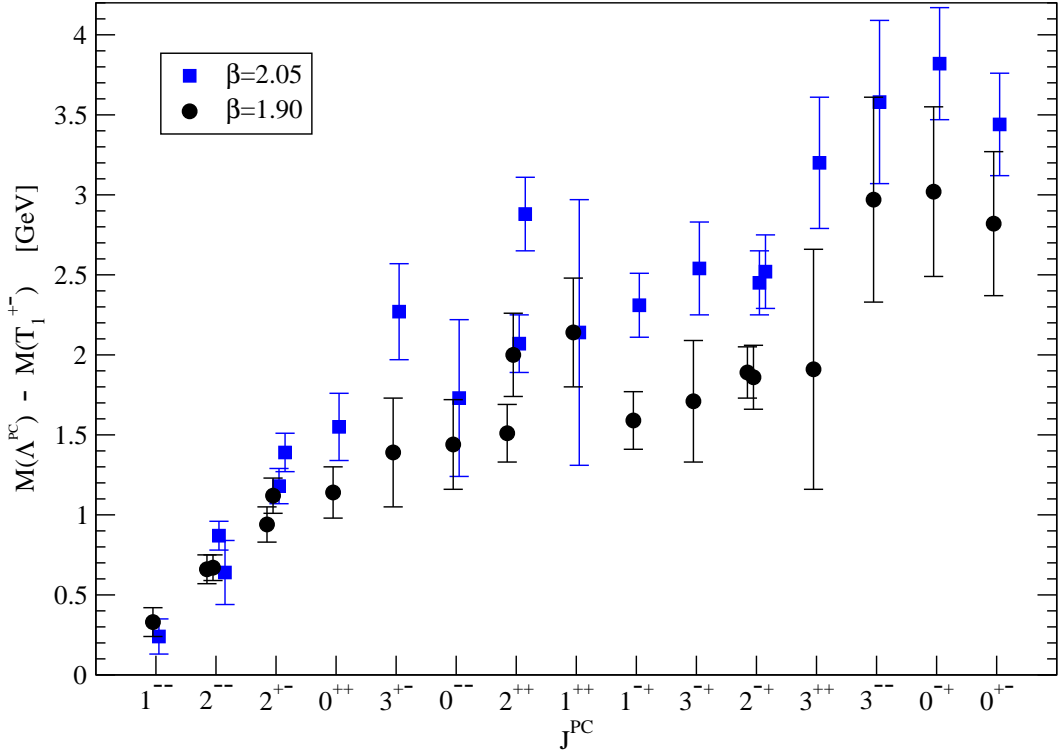}
\caption{The content of Table~\ref{table:OctetMass} is displayed visually.
Statistical and systematic errors were added linearly.}
\label{fig:octetspectrum} 
\end{figure}

\begin{table}[tb]
\caption{The mass spectrum of gluelumps containing a static octet particle, as
determined from dynamical lattice QCD at two lattice spacings.  $J$
denotes the continuum angular momentum of the light (gauge) degrees of freedom and does not include the spin of the octet particle.  
The first error is statistical and the second is
systematic, from Eq.~(\ref{systematicerror}).}
\label{table:OctetMass} 
\begin{tabular}{lccc}
\hline
\hline
$\Lambda^{PC}$ & ~~~~~$J$~~~~~
 & \multicolumn{2}{c}{$M(\Lambda^{PC})-M(T_1^{+-})$~~ [GeV]} \\
\cline{3-4}
 & & \multicolumn{1}{c}{~~~~~~$\beta=1.90$~~~~~~}
   & \multicolumn{1}{c}{~~~~~~$\beta=2.05$~~~~~~} \\
\hline
$T_1^{--}$ & 1 & 0.33$\pm$0.02$\pm$0.07 & 0.24$\pm$0.05$\pm$0.06 \\
$E^{--}  $ & 2 & 0.66$\pm$0.02$\pm$0.07 & 0.87$\pm$0.04$\pm$0.05 \\
$T_2^{--}$ & 2 & 0.67$\pm$0.02$\pm$0.06 & 0.64$\pm$0.11$\pm$0.09 \\
$E^{+-}  $ & 2 & 0.94$\pm$0.03$\pm$0.08 & 1.18$\pm$0.05$\pm$0.06 \\
$T_2^{+-}$ & 2 & 1.12$\pm$0.03$\pm$0.08 & 1.39$\pm$0.06$\pm$0.06 \\
$A_1^{++}$ & 0 & 1.14$\pm$0.05$\pm$0.11 & 1.55$\pm$0.12$\pm$0.09 \\
$A_2^{+-}$ & 3 & 1.39$\pm$0.12$\pm$0.22 & 2.27$\pm$0.05$\pm$0.25 \\
$A_1^{--}$ & 0 & 1.44$\pm$0.09$\pm$0.19 & 1.73$\pm$0.23$\pm$0.26 \\
$E^{++}  $ & 2 & 1.51$\pm$0.07$\pm$0.11 & 2.07$\pm$0.03$\pm$0.15 \\
$T_2^{++}$ & 2 & 2.00$\pm$0.13$\pm$0.13 & 2.88$\pm$0.05$\pm$0.18 \\
$T_1^{++}$ & 1 & 2.14$\pm$0.15$\pm$0.19 & 2.14$\pm$0.38$\pm$0.45 \\
$T_1^{-+}$ & 1 & 1.59$\pm$0.06$\pm$0.12 & 2.31$\pm$0.04$\pm$0.16 \\
$A_2^{-+}$ & 3 & 1.71$\pm$0.14$\pm$0.24 & 2.54$\pm$0.06$\pm$0.23 \\
$E^{-+}  $ & 2 & 1.89$\pm$0.10$\pm$0.06 & 2.45$\pm$0.04$\pm$0.16 \\
$T_2^{-+}$ & 2 & 1.86$\pm$0.09$\pm$0.11 & 2.52$\pm$0.04$\pm$0.19 \\
$A_2^{++}$ & 3 & 1.91$\pm$0.33$\pm$0.42 & 3.20$\pm$0.12$\pm$0.29 \\
$A_2^{--}$ & 3 & 2.97$\pm$0.05$\pm$0.59 & 3.58$\pm$0.19$\pm$0.32 \\
$A_1^{-+}$ & 0 & 3.02$\pm$0.05$\pm$0.48 & 3.82$\pm$0.18$\pm$0.17 \\
$A_1^{+-}$ & 0 & 2.82$\pm$0.04$\pm$0.41 & 3.44$\pm$0.13$\pm$0.19 \\
\hline
\hline
\end{tabular}
\end{table}
Table~\ref{table:OctetMass} and Fig.~\ref{fig:octetspectrum}
contain the final results for mass splittings
among gluelumps with the static particle in the color-octet representation.
As is true throughout this article, angular momentum $J$ refers to the light
degrees of freedom only;
all results apply to a heavy particle---color octet in this case---with {\em any} spin.
Although the central value for the mass difference
at $\beta=2.05$ tends to be larger than the central value at $\beta=1.90$,
the effect is marginal relative to the quoted error bars.
Since both lattice spacings are less than 0.1 fm and an improved lattice QCD
action has been used, it is not surprising that mass splittings are
essentially independent of lattice spacing.  It is also reassuring to see
that $E^{PC}$ and $T_2^{PC}$ are consistent with each other for each $PC$,
since they should couple to the same physical state ($J$ = 2) in the continuum
limit.

The quenched lattice QCD study of
Ref.~\cite{Foster:1998wu} had access to only 10 of the 20 channels listed in
our Table~\ref{table:OctetMass}. The raw data for those 10 channels are
provided in Table~II of Ref.~\cite{Foster:1998wu} (here called
``\cite{Foster:1998wu}-II'' for brevity) without systematic errors,
but several options for adjacent time steps are shown in \cite{Foster:1998wu}-II
and from this a systematic error defined by our
Eq.~(\ref{systematicerror}) can be estimated if desired.
The raw data from \cite{Foster:1998wu}-II
are in reasonable agreement with the present work, but we wish to point out
some concerns about how \cite{Foster:1998wu}-II was used to
arrive at final mass splittings in MeV, as listed in \cite{Foster:1998wu}-III.

To begin, we note that \cite{Foster:1998wu}-III was obtained from
\cite{Foster:1998wu}-II by going through the figure \cite{Foster:1998wu}-3.
The figure \cite{Foster:1998wu}-3 is largely obtained
from \cite{Foster:1998wu}-II by using the first two time steps
(called ``$t$=2:1'' in Ref.~\cite{Foster:1998wu})
and combining errors from the two energy levels in quadrature.
This numerically reproduces the data in \cite{Foster:1998wu}-3
with two exceptions, both at $\beta=5.7$:
the $A_1^{++}$ data point in \cite{Foster:1998wu}-3
is not consistent with
\cite{Foster:1998wu}-II, and neither is the error bar for $T_2^{++}$.
A first concern is that $\beta=5.7$ data have a significant impact on the
continuum extrapolation presented in \cite{Foster:1998wu}-III, and direct use of
\cite{Foster:1998wu}-II indicates a much more modest lattice spacing dependence
than was claimed.
A second concern is that the $t$=2:1 data are used to obtain
the continuum limit even though \cite{Foster:1998wu}-II shows that they produce
mass splittings that differ significantly from later time steps.

These concerns should not detract from the valuable comparison between
the present study and \cite{Foster:1998wu}-II.
Our dynamical lattice QCD study uses two lattice spacings that are very close
to the finer two spacings of the quenched study in Ref.~\cite{Foster:1998wu},
and produces compatible results, which indicates that quenching errors are too
small to disentangle from the other uncertainties.
The authors of Ref.~\cite{Foster:1998wu} reported a lack of
degeneracy for $E^{++}$ and $T_2^{++}$ at nonzero lattice spacings, with
$T_2^{++}$ heavier than $E^{++}$, and we see a similar tendency though it is
not large relative to the error bars in Fig.~\ref{fig:octetspectrum}.
Moreover, we now have three other channels ($+-$, $-+$, and $--$) where
$E$ and $T_2$ can be compared, and these are all
appropriately degenerate when systematic uncertainties are taken into account.
The authors of Ref.~\cite{Foster:1998wu} were surprised by the degeneracy
of $2^{+-}$ with $3^{+-}$, but in the context of our 20-channel study this
pair of operators has no striking degeneracy.
The authors of Ref.~\cite{Foster:1998wu} were surprised by the heaviness of
the $0^{++}$, and we agree that it is heavy, although the extrapolation in Ref.~\cite{Foster:1998wu} is
noticeably reduced when
the $\beta=5.7$ data are taken directly from \cite{Foster:1998wu}-II.

\begin{table}[tb]
\caption{The mass spectrum of gluelumps containing a static decuplet particle,
as determined from dynamical lattice QCD at $\beta=1.90$.  $J$
denotes the continuum angular momentum of the light (gauge) degrees of freedom and does not include the spin of the decuplet particle.  
The first error is statistical and the second is
systematic, from Eq.~(\ref{systematicerror}).}
\label{table:DecupletMass} 
\begin{tabular}{lcc}
\hline
\hline
$\Lambda^P$ & ~~~~~$J$~~~~~ & $M(\Lambda^P)-M(A_1^-)$~~ [GeV] \\
\hline
$T_1^-$ & 1 & 0.39$\pm$0.04$\pm$0.33 \\
$E^-  $ & 2 & 0.40$\pm$0.05$\pm$0.33 \\
$T_2^-$ & 2 & 0.41$\pm$0.04$\pm$0.32 \\
$T_1^+$ & 1 & 0.57$\pm$0.05$\pm$0.48 \\
$A_2^+$ & 3 & 0.89$\pm$0.10$\pm$0.47 \\
$E^+  $ & 2 & 0.90$\pm$0.07$\pm$0.45 \\
$T_2^+$ & 2 & 0.96$\pm$0.06$\pm$0.37 \\
$A_1^+$ & 0 & 1.05$\pm$0.10$\pm$0.38 \\
$A_2^-$ & 3 & 1.48$\pm$0.17$\pm$0.44 \\
\hline
\hline
\end{tabular}
\end{table}
For representations beyond the octet, our efforts to optimize the smeared operators were concentrated on the $\beta=1.90$ ensemble.  Simulations of the $\beta=2.05$ lattices were computationally expensive and, like the octet results, we do not anticipate a significant
dependence on lattice spacing between these two $\beta$ values, so our
results at $\beta=1.90$ represent predictions for the
continuum physics spectrum.
Mass splittings for gluelumps with the static particle in the
color-decuplet representation are shown in Table~\ref{table:DecupletMass}.
Notice that the $A_1^-$, which in the continuum is $0^-$, appears to be the
lightest state in this spectrum modulo systematic uncertainties.

Mass splittings for gluelumps containing a 27-plet static particle are shown in
Table~\ref{table:27PletMass}.  All 20 $\Lambda^{PC}$ channels were attempted,
but those omitted from the table produced no usable signal.
Although mass differences are tabulated relative to $T_2^{++}$, the data
do not ensure that this is the lightest state.
For both the decuplet and the 27-plet, the $E$ and $T_2$ channels are consistent with one another.
\begin{table}[tb]
\caption{The resolvable mass spectrum of gluelumps containing a static 27-plet particle,
as determined from dynamical lattice QCD at $\beta=1.90$.  $J$
denotes the continuum angular momentum of the light (gauge) degrees of freedom and does not include the spin of the 27-plet particle.  
The first error is statistical and the second is
systematic, from Eq.~(\ref{systematicerror}).}
\label{table:27PletMass} 
\begin{tabular}{lcc}
\hline
\hline
$\Lambda^{PC}$ & ~~~~~$J$~~~~~ & $M(\Lambda^{PC})-M(T_2^{++})$~~ [GeV] \\
\hline
$E^{++}$   & 2 &  0.04$\pm$0.06$\pm$0.24 \\
$A_1^{++}$ & 0 &  0.05$\pm$0.07$\pm$0.19 \\
$T_2^{-+}$ & 2 &  0.50$\pm$0.08$\pm$0.45 \\
$E^{-+}$   & 2 &  0.53$\pm$0.09$\pm$0.43 \\
$T_1^{-+}$ & 1 &  0.66$\pm$0.09$\pm$0.38 \\
$A_2^{-+}$ & 3 &  0.74$\pm$0.17$\pm$0.50 \\
$A_2^{++}$ & 3 &  1.12$\pm$0.23$\pm$0.45 \\
$T_1^{+-}$ & 1 &  1.29$\pm$0.26$\pm$0.93 \\
$T_1^{++}$ & 1 &  1.34$\pm$0.20$\pm$0.57 \\
$T_2^{+-}$ & 2 &  1.91$\pm$0.32$\pm$0.45 \\
$E^{+-}  $ & 2 &  2.13$\pm$0.41$\pm$0.30 \\
$T_1^{--}$ & 1 &  2.49$\pm$0.55$\pm$0.44 \\
\hline
\hline
\end{tabular}
\end{table}

Correlation functions for the 28-plet and 35-plet contained too few usable
time steps to give a meaningful systematic error, so we refrain from
presenting numerical results.
Nevertheless, the writing and running of this code helped us to
confirm the operator definitions
presented in Secs.~\ref{sec:propagators} and \ref{sec:operators} and
Appendixes~\ref{app:tensors} and \ref{app:operators}---for example, 
we tested gauge invariance through explicit computations with
a single configuration in every case.

Although none of the operators used in the present work contain valence
quarks, physical states with valence quarks could have the same
quantum numbers as gluelumps.
Examples of such states include the adjoint mesons in
Ref.~\cite{Foster:1998wu} that were
explored on quenched lattices by using operators
that contain explicit valence quarks.
As exemplified by Fig.~8 of Ref.~\cite{Foster:1998wu}, the mass difference
between gluelumps and adjoint mesons is difficult to ascertain.
Our use of dynamical configurations in principle allows adjoint mesons to
mix with the gluelump signals, but our exclusive use of quark-free
operators likely produces only a feeble coupling to adjoint mesons.
A combined study of adjoint mesons and gluelumps would require operators
of both types to be analyzed simultaneously in a matrix that permits mixing
between them.

\section{Conclusions}

Any extension of the standard model with a long-lived colored heavy particle
will contain new hadrons that are QCD bound states of the heavy particle
together with gluons and quarks.
The lattice QCD study of this new hadron spectrum was
pioneered by Michael and collaborators
\cite{Campbell:1985kp,Jorysz:1987qj,Michael:1991nc,Foster:1998wu}, motivated
by the color-octet gluino of supersymmetry.

The present study has revisited the gluelump spectrum in greater detail.
This is the first lattice simulation to explore the complete set of gluelump
quantum numbers, $J^{PC}$, where $J$ represents the angular momentum of the
light degrees of freedom.  The heavy particle is treated as static, so its
spin decouples.
The lightest new state not studied previously is $0^{--}$,
which is found to be as light as some of the states that were studied in
Ref.~\cite{Foster:1998wu}.
Comparison of $E$ and $T_2$ representations, both of which couple to $J=2$ in
the continuum limit, provides a cross-check on systematic errors.
A leading systematic error was identified as arising from the choice of a
fitting window in Euclidean time.
Comparison of the quenched results from Ref.~\cite{Foster:1998wu} with the
present dynamical results does not reveal any large quenching artifacts.

In addition,
the present study provides the first results for generalized gluelumps, where
the heavy particle is not color octet but rather decuplet or 27-plet.
The machinery for 28-plet and 35-plet computations was also established and
tested, so future studies will be straightforward in those cases as well.

Final numerical results are presented in
Tables~\ref{table:OctetMass}, \ref{table:DecupletMass},
and \ref{table:27PletMass}.
The two $\beta$ values for octet results represent two different lattice
spacings that agree within uncertainties.
Comparison of Table~\ref{table:OctetMass} with the previous studies tabulated
in Table~\ref{table:FMresults} shows a general agreement, and indicates
that systematic errors cannot be neglected:
lattice results are presently limited by systematics rather than statistics.
Future studies can directly use the operators developed here to perform
larger-scale simulations and improve the precision for this spectrum of
generalized gluelumps.

\section*{Acknowledgments}
We thank the CP-PACS and JLQCD Collaborations for making their dynamical gauge
field configurations available. This work was supported in part by the Natural
Sciences and Engineering Research Council (NSERC) of Canada and by Compute
Canada through the Shared Hierarchical Academic Research Computing Network
(SHARCNET).

\appendix
\section{BASIS TENSORS FOR EACH REPRESENTATION}\label{app:tensors}

To reduce notational clutter, define generalized Kronecker delta functions
where indices in parenthesis are to be permuted through all distinct orderings.
A few examples are the following:
\begin{eqnarray}
\delta_{\{ij\}\{kk\}} & = & \delta_{ik}\delta_{jk} \,,
\nonumber \\
\delta_{\{ij\}\{kl\}} & = & \delta_{ik}\delta_{jl} + \delta_{il}\delta_{jk} \,,
\nonumber \\
\delta_{\{ijk\}\{lll\}} & = & \delta_{il}\delta_{jl}\delta_{kl} \,,
\nonumber \\
\delta_{\{ijk\}\{llm\}} & = & \delta_{il}\delta_{jl}\delta_{km} + \delta_{il}\delta_{jm}\delta_{kl} + \delta_{im}\delta_{jl}\delta_{kl} \,,
\nonumber \\
\delta_{\{ijk\}\{lmn\}} & = & \delta_{il}\delta_{jm}\delta_{kn} + \delta_{il}\delta_{jn}\delta_{km} + \delta_{im}\delta_{jl}\delta_{kn} + \delta_{im}\delta_{jn}\delta_{kl} + \delta_{in}\delta_{jl}\delta_{km} + \delta_{in}\delta_{jm}\delta_{kl} \,, \nonumber \\
\delta_{\{ijkl\}\{pppp\}} & = & \delta_{ip}\delta_{jp}\delta_{kp}\delta_{lp} \,,
\nonumber \\
\delta_{\{ijkl\}\{pppq\}} & = & \delta_{ip}\delta_{jp}\delta_{kp}\delta_{lq}
+ \delta_{ip}\delta_{jp}\delta_{kq}\delta_{lp}
+ \delta_{ip}\delta_{jq}\delta_{kp}\delta_{lp}
+ \delta_{iq}\delta_{jp}\delta_{kp}\delta_{lp} \,,
\nonumber \\
\delta_{\{ijkl\}\{ppqq\}} & = & \delta_{ip}\delta_{jp}\delta_{kq}\delta_{lq}
+ \delta_{ip}\delta_{jq}\delta_{kp}\delta_{lq}
+ \delta_{iq}\delta_{jp}\delta_{kp}\delta_{lq} \nonumber \\ & &
+ \delta_{ip}\delta_{jq}\delta_{kq}\delta_{lp}
+ \delta_{iq}\delta_{jp}\delta_{kq}\delta_{lp}
+ \delta_{iq}\delta_{jq}\delta_{kp}\delta_{lp} \,,
\nonumber \\
\delta_{\{ijkl\}\{ppqr\}} & = & \delta_{ip}\delta_{jp}\delta_{kq}\delta_{lr}
+ \delta_{ip}\delta_{jq}\delta_{kp}\delta_{lr}
+ \delta_{iq}\delta_{jp}\delta_{kp}\delta_{lr} \nonumber \\ &&
+ \delta_{ip}\delta_{jq}\delta_{kr}\delta_{lp}
+ \delta_{iq}\delta_{jp}\delta_{kr}\delta_{lp}
+ \delta_{iq}\delta_{jr}\delta_{kp}\delta_{lp} \nonumber \\ &&
+ \delta_{ip}\delta_{jp}\delta_{kr}\delta_{lq}
+ \delta_{ip}\delta_{jr}\delta_{kp}\delta_{lq}
+ \delta_{ir}\delta_{jp}\delta_{kp}\delta_{lq} \nonumber \\ &&
+ \delta_{ip}\delta_{jr}\delta_{kq}\delta_{lp}
+ \delta_{ir}\delta_{jp}\delta_{kq}\delta_{lp}
+ \delta_{ir}\delta_{jq}\delta_{kp}\delta_{lp} \,.
\end{eqnarray}
The basis used for the octet representation is
\begin{equation}
\begin{array}{l l l}
T^1_{ij} = \delta_{i1}\delta_{j2} \,, \qquad \qquad \qquad \quad &
T^2_{ij} = \delta_{i1}\delta_{j3} \,, \qquad \qquad \qquad \quad &
T^3_{ij} = \delta_{i2}\delta_{j3} \,, \\
T^4_{ij} = \delta_{i2}\delta_{j1} \,, &
T^5_{ij} = \delta_{i3}\delta_{j1} \,, &
T^6_{ij} = \delta_{i3}\delta_{j2} \,, \\
T^7_{ij} = \frac{1}{2} \left(\delta_{i1}\delta_{j1} -\delta_{i2}\delta_{j2} \right) \,, &
\multicolumn{2}{l}{T^8_{ij} = \frac{1}{\sqrt{6}} \left(\delta_{i1}\delta_{j1} + \delta_{i2}\delta_{j2} -2\delta_{i3}\delta_{j3} \right)} \,.
\end{array}
\end{equation}
The octet tensors obey the relation
\begin{equation}
\sum_{\mu = 1}^{8} T^{\mu}_{ij} T^{\mu}_{kl} = \delta_{ik} \delta_{jl} - \frac{1}{3} \delta_{ij} \delta_{kl} \, .
\end{equation}
The basis used for the decuplet representation is
\begin{equation}
\begin{array}{l l l}
 T^1_{ijk} = \delta_{\{ijk\}\{111\}} \,, 
& T^2_{ijk} = \delta_{\{ijk\}\{222\}} \,, 
& T^3_{ijk} = \delta_{\{ijk\}\{333\}} \,, 
 \\
 T^4_{ijk} = \frac{1}{\sqrt{3}} \delta_{\{ijk\}\{112\}} \,, \qquad \qquad
& T^5_{ijk} = \frac{1}{\sqrt{3}} \delta_{\{ijk\}\{113\}} \,, \qquad \qquad
& T^6_{ijk} = \frac{1}{\sqrt{3}} \delta_{\{ijk\}\{122\}} \,, 
 \\
 T^7_{ijk} = \frac{1}{\sqrt{3}} \delta_{\{ijk\}\{223\}} \,, 
& T^8_{ijk} = \frac{1}{\sqrt{3}} \delta_{\{ijk\}\{133\}} \,, 
& T^9_{ijk} = \frac{1}{\sqrt{3}} \delta_{\{ijk\}\{233\}} \,, 
 \\
T^{10}_{ijk} = \frac{1}{\sqrt{6}} \delta_{\{ijk\}\{123\}} \,.
&& \\
\end{array}
\end{equation}
The decuplet tensors obey the relation
\begin{equation}
\sum_{\mu = 1}^{10} T^{\mu}_{ijk} T^{\mu}_{lmn} = \frac{1}{6} \delta_{\{ijk\}} \delta_{\{lmn\}} \, .
\end{equation}
The basis used for the 27-plet representation is
\begin{eqnarray}
&&
 T^1_{ijkl} = \delta_{\{ij\}\{11\}} \delta_{\{kl\}\{22\}} \,, \qquad
 T^2_{ijkl} = \delta_{\{ij\}\{11\}} \delta_{\{kl\}\{33\}} \,, \qquad
 T^3_{ijkl} = \delta_{\{ij\}\{22\}} \delta_{\{kl\}\{33\}} \,, 
\nonumber \\
&&
 T^4_{ijkl} = \delta_{\{ij\}\{22\}} \delta_{\{kl\}\{11\}} \,, \qquad
 T^5_{ijkl} = \delta_{\{ij\}\{33\}} \delta_{\{kl\}\{11\}} \,, \qquad
 T^6_{ijkl} = \delta_{\{ij\}\{33\}} \delta_{\{kl\}\{22\}} \,, 
\nonumber \\
&&
 T^7_{ijkl} = \frac{1}{\sqrt{2}} \left( \delta_{\{ij\}\{11\}} \delta_{\{kl\}\{23\}} \right) \,, \qquad
 T^8_{ijkl} = \frac{1}{\sqrt{2}} \left( \delta_{\{ij\}\{12\}} \delta_{\{kl\}\{33\}} \right) \,, 
\nonumber \\
&&
 T^9_{ijkl} = \frac{1}{\sqrt{2}} \left( \delta_{\{ij\}\{13\}} \delta_{\{kl\}\{22\}} \right) \,, \qquad
 T^{10}_{ijkl} = \frac{1}{\sqrt{2}} \left( \delta_{\{ij\}\{23\}} \delta_{\{kl\}\{11\}} \right) \,, 
\nonumber \\
&&
 T^{11}_{ijkl} = \frac{1}{\sqrt{2}} \left( \delta_{\{ij\}\{33\}} \delta_{\{kl\}\{12\}} \right) \,, \qquad
 T^{12}_{ijkl} = \frac{1}{\sqrt{2}} \left( \delta_{\{ij\}\{22\}} \delta_{\{kl\}\{13\}} \right) \,, 
\nonumber \\
&&
 T^{13}_{ijkl} = \frac{1}{2} \left( \delta_{\{ij\}\{11\}} \delta_{\{kl\}\{12\}} -\delta_{\{ij\}\{12\}} \delta_{\{kl\}\{22\}}  \right) \,,
\nonumber \\
&&
 T^{14}_{ijkl} = \frac{1}{\sqrt{20}} \left( \delta_{\{ij\}\{11\}} \delta_{\{kl\}\{12\}} +\delta_{\{ij\}\{12\}} \delta_{\{kl\}\{22\}} -2\delta_{\{ij\}\{13\}} \delta_{\{kl\}\{23\}} \right) \,,
\nonumber \\
&&
 T^{15}_{ijkl} = \frac{1}{2} \left( \delta_{\{ij\}\{11\}} \delta_{\{kl\}\{13\}} -\delta_{\{ij\}\{13\}} \delta_{\{kl\}\{33\}}  \right) \,, 
\nonumber \\
&&
 T^{16}_{ijkl} = \frac{1}{\sqrt{20}} \left( \delta_{\{ij\}\{11\}} \delta_{\{kl\}\{13\}} +\delta_{\{ij\}\{13\}} \delta_{\{kl\}\{33\}} -2\delta_{\{ij\}\{12\}} \delta_{\{kl\}\{23\}}  \right) \,, 
\nonumber \\
&&
 T^{17}_{ijkl} = \frac{1}{2} \left( \delta_{\{ij\}\{12\}} \delta_{\{kl\}\{11\}} -\delta_{\{ij\}\{22\}} \delta_{\{kl\}\{12\}}  \right) \,, 
\nonumber \\
&&
 T^{18}_{ijkl} = \frac{1}{\sqrt{20}} \left( \delta_{\{ij\}\{12\}} \delta_{\{kl\}\{11\}} +\delta_{\{ij\}\{22\}} \delta_{\{kl\}\{12\}} -2\delta_{\{ij\}\{23\}} \delta_{\{kl\}\{13\}}  \right) \,,
\nonumber \\
&&
 T^{19}_{ijkl} = \frac{1}{2} \left( \delta_{\{ij\}\{22\}} \delta_{\{kl\}\{23\}} -\delta_{\{ij\}\{23\}} \delta_{\{kl\}\{33\}}  \right) \,, 
\nonumber \\
&&
 T^{20}_{ijkl} = \frac{1}{\sqrt{20}} \left( \delta_{\{ij\}\{22\}} \delta_{\{kl\}\{23\}} +\delta_{\{ij\}\{23\}} \delta_{\{kl\}\{33\}} -2\delta_{\{ij\}\{12\}} \delta_{\{kl\}\{13\}} \right) \,, 
\nonumber \\
&&
 T^{21}_{ijkl} = \frac{1}{2} \left( \delta_{\{ij\}\{13\}} \delta_{\{kl\}\{11\}} -\delta_{\{ij\}\{33\}} \delta_{\{kl\}\{13\}}  \right) \,, 
\nonumber \\
&&
 T^{22}_{ijkl} = \frac{1}{\sqrt{20}} \left( \delta_{\{ij\}\{13\}} \delta_{\{kl\}\{11\}} +\delta_{\{ij\}\{33\}} \delta_{\{kl\}\{13\}} -2\delta_{\{ij\}\{23\}} \delta_{\{kl\}\{12\}} \right) \,, 
\nonumber \\
&&
 T^{23}_{ijkl} = \frac{1}{2} \left( \delta_{\{ij\}\{23\}} \delta_{\{kl\}\{22\}} -\delta_{\{ij\}\{33\}} \delta_{\{kl\}\{23\}} \right) \,, 
\nonumber \\
&&
 T^{24}_{ijkl} = \frac{1}{\sqrt{20}} \left( \delta_{\{ij\}\{23\}} \delta_{\{kl\}\{22\}} +\delta_{\{ij\}\{33\}} \delta_{\{kl\}\{23\}} -2\delta_{\{ij\}\{13\}} \delta_{\{kl\}\{12\}} \right) \,, 
\nonumber \\
&&
 T^{25}_{ijkl} = \frac{1}{\sqrt{10}} \left( \delta_{\{ij\}\{11\}} \delta_{\{kl\}\{11\}} - \delta_{\{ij\}\{22\}} \delta_{\{kl\}\{22\}} - \delta_{\{ij\}\{13\}} \delta_{\{kl\}\{13\}} + \delta_{\{ij\}\{23\}} \delta_{\{kl\}\{23\}} \right) \,, 
\nonumber \\
&&
 T^{26}_{ijkl} = \frac{1}{\sqrt{30}}  \left( \delta_{\{ij\}\{11\}} \delta_{\{kl\}\{11\}} + \delta_{\{ij\}\{22\}} \delta_{\{kl\}\{22\}} - 2\delta_{\{ij\}\{33\}} \delta_{\{kl\}\{33\}} \right. 
\nonumber \\
&&              \qquad \qquad         \left. - 2\delta_{\{ij\}\{12\}} \delta_{\{kl\}\{12\}} + \delta_{\{ij\}\{13\}} \delta_{\{kl\}\{13\}} + \delta_{\{ij\}\{23\}} \delta_{\{kl\}\{23\}} \right)  \,,
\nonumber \\
&&
 T^{27}_{ijkl} = \frac{1}{\sqrt{24}}  \left( 2\delta_{\{ij\}\{11\}} \delta_{\{kl\}\{11\}} + 2\delta_{\{ij\}\{22\}} \delta_{\{kl\}\{22\}} + 2\delta_{\{ij\}\{33\}} \delta_{\{kl\}\{33\}} \right. 
\nonumber \\
&&              \qquad \qquad         \left. - \delta_{\{ij\}\{12\}} \delta_{\{kl\}\{12\}} - \delta_{\{ij\}\{13\}} \delta_{\{kl\}\{13\}} - \delta_{\{ij\}\{23\}} \delta_{\{kl\}\{23\}} \right) \,.
\end{eqnarray}
The 27-plet tensors obey the relation
\begin{eqnarray}
\sum_{\mu = 1}^{27} T^{\mu}_{ijkl} T^{\mu}_{mnop} & = 
& \frac{1}{4} \left( \delta_{im}\delta_{jn}\delta_{ko}\delta_{lp} + \delta_{im}\delta_{jn}\delta_{kp}\delta_{lo} \right. 
\nonumber \\
&&            \left. \quad + \delta_{in}\delta_{jm}\delta_{ko}\delta_{lp} + \delta_{in}\delta_{jm}\delta_{kp}\delta_{lo} \right) 
\nonumber \\
&& -\frac{1}{20} \left( \delta_{im}\delta_{jl}\delta_{ko}\delta_{np} + \delta_{im}\delta_{jl}\delta_{kp}\delta_{no} \right. 
\nonumber \\
&& \qquad           + \delta_{in}\delta_{jl}\delta_{ko}\delta_{mp} + \delta_{in}\delta_{jl}\delta_{kp}\delta_{mo}  
\nonumber \\
&& \qquad           + \delta_{im}\delta_{lo}\delta_{jk}\delta_{np} + \delta_{im}\delta_{lp}\delta_{jk}\delta_{no}  
\nonumber \\
&& \qquad           + \delta_{in}\delta_{lo}\delta_{jk}\delta_{mp} + \delta_{in}\delta_{lp}\delta_{jk}\delta_{mo} 
\nonumber \\
&& \qquad           + \delta_{il}\delta_{jm}\delta_{ko}\delta_{np} + \delta_{il}\delta_{jm}\delta_{kp}\delta_{no} 
\nonumber \\
&& \qquad           + \delta_{il}\delta_{jn}\delta_{ko}\delta_{mp} + \delta_{il}\delta_{jn}\delta_{kp}\delta_{mo} 
\nonumber \\
&& \qquad           + \delta_{ik}\delta_{jm}\delta_{lo}\delta_{np} + \delta_{ik}\delta_{jm}\delta_{lp}\delta_{no} 
\nonumber \\
&& \qquad           + \left. \delta_{ik}\delta_{jn}\delta_{lo}\delta_{mp} + \delta_{ik}\delta_{jn}\delta_{lp}\delta_{mo} \right) 
\nonumber \\
&& +\frac{1}{40} \left( \delta_{ik}\delta_{jl}\delta_{mo}\delta_{np} + \delta_{ik}\delta_{jl}\delta_{mp}\delta_{no} \right. 
\nonumber \\
&&            \left. \qquad+ \delta_{il}\delta_{jk}\delta_{mo}\delta_{np} + \delta_{il}\delta_{jk}\delta_{mp}\delta_{no} \right) \,.
\end{eqnarray}
The basis used for the 28-plet representation is
\begin{eqnarray}
& & T^{1}_{ijklmn} = \delta_{\{ijklmn\}\{111111\}} \,, \qquad \qquad \qquad
 T^{2}_{ijklmn} = \delta_{\{ijklmn\}\{222222\}} \,, \nonumber \\
& & T^{3}_{ijklmn} = \delta_{\{ijklmn\}\{333333\}} \,, \qquad \qquad \qquad 
 T^{4}_{ijklmn} = \frac{1}{\sqrt{6}} \delta_{\{ijklmn\}\{111112\}} \,, \nonumber \\
& & T^{5}_{ijklmn} = \frac{1}{\sqrt{6}} \delta_{\{ijklmn\}\{111113\}} \,, \qquad \qquad \; \,
 T^{6}_{ijklmn} = \frac{1}{\sqrt{6}} \delta_{\{ijklmn\}\{222221\}} \,, \nonumber \\
& & T^{7}_{ijklmn} = \frac{1}{\sqrt{6}} \delta_{\{ijklmn\}\{222223\}} \,, \qquad \qquad \; \,
 T^{8}_{ijklmn} = \frac{1}{\sqrt{6}} \delta_{\{ijklmn\}\{333331\}} \,, \nonumber \\
& & T^{9}_{ijklmn} = \frac{1}{\sqrt{6}} \delta_{\{ijklmn\}\{333332\}} \,, \qquad \qquad \; \,
 T^{10}_{ijklmn} = \frac{1}{\sqrt{15}} \delta_{\{ijklmn\}\{111122\}} \,, \nonumber \\
& & T^{11}_{ijklmn} = \frac{1}{\sqrt{15}} \delta_{\{ijklmn\}\{111133\}} \,, \qquad \qquad
 T^{12}_{ijklmn} = \frac{1}{\sqrt{15}} \delta_{\{ijklmn\}\{222211\}} \,, \nonumber \\
& & T^{13}_{ijklmn} = \frac{1}{\sqrt{15}} \delta_{\{ijklmn\}\{222233\}} \,, \qquad \qquad
 T^{14}_{ijklmn} = \frac{1}{\sqrt{15}} \delta_{\{ijklmn\}\{333311\}} \,, \nonumber \\
& & T^{15}_{ijklmn} = \frac{1}{\sqrt{15}} \delta_{\{ijklmn\}\{333322\}} \,, \qquad \qquad 
 T^{16}_{ijklmn} = \frac{1}{\sqrt{20}} \delta_{\{ijklmn\}\{111222\}} \,, \nonumber \\
& & T^{17}_{ijklmn} = \frac{1}{\sqrt{20}} \delta_{\{ijklmn\}\{111333\}} \,, \qquad \qquad
 T^{18}_{ijklmn} = \frac{1}{\sqrt{20}} \delta_{\{ijklmn\}\{222333\}} \,, \nonumber \\
& & T^{19}_{ijklmn} = \frac{1}{\sqrt{30}} \delta_{\{ijklmn\}\{111123\}} \,, \qquad \qquad
 T^{20}_{ijklmn} = \frac{1}{\sqrt{30}} \delta_{\{ijklmn\}\{222213\}} \,, \nonumber \\
& & T^{21}_{ijklmn} = \frac{1}{\sqrt{30}} \delta_{\{ijklmn\}\{333312\}} \,, \qquad \qquad 
 T^{22}_{ijklmn} = \frac{1}{\sqrt{60}} \delta_{\{ijklmn\}\{111223\}} \,, \nonumber \\
& & T^{23}_{ijklmn} = \frac{1}{\sqrt{60}} \delta_{\{ijklmn\}\{111332\}} \,, \qquad \qquad
 T^{24}_{ijklmn} = \frac{1}{\sqrt{60}} \delta_{\{ijklmn\}\{222113\}} \,, \nonumber \\
& & T^{25}_{ijklmn} = \frac{1}{\sqrt{60}} \delta_{\{ijklmn\}\{222331\}} \,, \qquad \qquad
 T^{26}_{ijklmn} = \frac{1}{\sqrt{60}} \delta_{\{ijklmn\}\{333112\}} \,, \nonumber \\
& & T^{27}_{ijklmn} = \frac{1}{\sqrt{60}} \delta_{\{ijklmn\}\{333221\}} \,, \qquad \qquad 
 T^{28}_{ijklmn} = \frac{1}{\sqrt{90}} \delta_{\{ijklmn\}\{112233\}} \,. 
\end{eqnarray}
The 28-plet tensors obey the relation
\begin{equation}
\sum_{\mu = 1}^{28} T^{\mu}_{ijklmn} T^{\mu}_{opqrst} = \frac{1}{6!} \delta_{\{ijklmn\}} \delta_{\{opqrst\}} \, .
\end{equation}
The basis used for the 35-plet representation is
\begin{eqnarray}
& & T^{1}_{ijklm} = \delta_{\{ijkl\}\{1111\}} \delta_{m3} \,, \qquad \qquad \quad \; \;
 T^{2}_{ijklm} = \delta_{\{ijkl\}\{1111\}} \delta_{m2} \,, \nonumber \\
& & T^{3}_{ijklm} = \delta_{\{ijkl\}\{2222\}} \delta_{m3} \,, \qquad \qquad \quad \; \;
 T^{4}_{ijklm} = \delta_{\{ijkl\}\{2222\}} \delta_{m1} \,, \nonumber \\
& & T^{5}_{ijklm} = \delta_{\{ijkl\}\{3333\}} \delta_{m2} \,, \qquad \qquad \quad \; \; 
 T^{6}_{ijklm} = \delta_{\{ijkl\}\{3333\}} \delta_{m1} \,, \nonumber \\
& & T^{7}_{ijklm} = \frac{1}{2} \delta_{\{ijkl\}\{1112\}} \delta_{m3} \,, \qquad \qquad \; \; \;
 T^{8}_{ijklm} = \frac{1}{2} \delta_{\{ijkl\}\{1113\}} \delta_{m2} \,, \nonumber \\
& & T^{9}_{ijklm} = \frac{1}{2} \delta_{\{ijkl\}\{1222\}} \delta_{m3} \,, \qquad \qquad \; \; \;
 T^{10}_{ijklm} = \frac{1}{2} \delta_{\{ijkl\}\{2223\}} \delta_{m1} \,, \nonumber \\
& & T^{11}_{ijklm} = \frac{1}{2} \delta_{\{ijkl\}\{1333\}} \delta_{m2} \,, \qquad \qquad \; \; \;
 T^{12}_{ijklm} = \frac{1}{2} \delta_{\{ijkl\}\{2333\}} \delta_{m1} \,, \nonumber \\
& & T^{13}_{ijklm} = \frac{1}{\sqrt{6}} \delta_{\{ijkl\}\{1122\}} \delta_{m3} \,, \qquad \qquad 
 T^{14}_{ijklm} = \frac{1}{\sqrt{6}} \delta_{\{ijkl\}\{1133\}} \delta_{m2} \,, \nonumber \\
& & T^{15}_{ijklm} = \frac{1}{\sqrt{6}} \delta_{\{ijkl\}\{2233\}} \delta_{m1} \,, \nonumber \\
& & T^{16}_{ijklm} = \frac{1}{\sqrt{24}} \left( \delta_{\{ijkl\}\{1231\}} \delta_{m1} - \delta_{\{ijkl\}\{1232\}} \delta_{m2} \right) \,, \nonumber \\
& & T^{17}_{ijklm} = \frac{1}{\sqrt{72}} \left( \delta_{\{ijkl\}\{1231\}} \delta_{m1} + \delta_{\{ijkl\}\{1232\}} \delta_{m2}  
-2 \delta_{\{ijkl\}\{1233\}} \delta_{m3} \right) \,, \nonumber \\
& & T^{18}_{ijklm} = \frac{1}{\sqrt{8}} \left( \delta_{\{ijkl\}\{1112\}} \delta_{m2} - \delta_{\{ijkl\}\{1113\}} \delta_{m3} \right) \,, \nonumber \\
& & T^{19}_{ijklm} = \frac{1}{\sqrt{12}} \left( \delta_{\{ijkl\}\{1112\}} \delta_{m2} + \delta_{\{ijkl\}\{1113\}} \delta_{m3}  
-2 \delta_{\{ijkl\}\{1111\}} \delta_{m1} \right) \,, \nonumber \\
& & T^{20}_{ijklm} = \frac{1}{\sqrt{8}} \left( \delta_{\{ijkl\}\{2221\}} \delta_{m1} - \delta_{\{ijkl\}\{2223\}} \delta_{m3} \right) \,, \nonumber \\
& & T^{21}_{ijklm} = \frac{1}{\sqrt{12}} \left( \delta_{\{ijkl\}\{2221\}} \delta_{m1} + \delta_{\{ijkl\}\{2223\}} \delta_{m3}  
-2 \delta_{\{ijkl\}\{2222\}} \delta_{m2} \right) \,, \nonumber \\
& & T^{22}_{ijklm} = \frac{1}{\sqrt{8}} \left( \delta_{\{ijkl\}\{3331\}} \delta_{m1} - \delta_{\{ijkl\}\{3332\}} \delta_{m2} \right) \,, \nonumber \\
& & T^{23}_{ijklm} = \frac{1}{\sqrt{12}} \left( \delta_{\{ijkl\}\{3331\}} \delta_{m1} + \delta_{\{ijkl\}\{3332\}} \delta_{m2}  
-2 \delta_{\{ijkl\}\{3333\}} \delta_{m3} \right) \,, \nonumber \\
& & T^{24}_{ijklm} = \frac{1}{\sqrt{18}} \left( \delta_{\{ijkl\}\{1122\}} \delta_{m2} - \delta_{\{ijkl\}\{1123\}} \delta_{m3} \right) \,, \nonumber \\
& & T^{25}_{ijklm} = \frac{1}{\sqrt{72}} \left(3\delta_{\{ijkl\}\{1121\}} \delta_{m1} -2\delta_{\{ijkl\}\{1122\}} \delta_{m2}  
-1 \delta_{\{ijkl\}\{1123\}} \delta_{m3} \right) \,, \nonumber \\
& & T^{26}_{ijklm} = \frac{1}{\sqrt{18}} \left( \delta_{\{ijkl\}\{1133\}} \delta_{m3} - \delta_{\{ijkl\}\{1132\}} \delta_{m2} \right) \,, \nonumber \\
& & T^{27}_{ijklm} = \frac{1}{\sqrt{72}} \left(3\delta_{\{ijkl\}\{1131\}} \delta_{m1} -2\delta_{\{ijkl\}\{1133\}} \delta_{m3}  
-1 \delta_{\{ijkl\}\{1132\}} \delta_{m2} \right) \,, \nonumber \\
& & T^{28}_{ijklm} = \frac{1}{\sqrt{18}} \left( \delta_{\{ijkl\}\{2211\}} \delta_{m1} - \delta_{\{ijkl\}\{2213\}} \delta_{m3} \right) \,, \nonumber \\
& & T^{29}_{ijklm} = \frac{1}{\sqrt{72}} \left(3\delta_{\{ijkl\}\{2212\}} \delta_{m2} -2\delta_{\{ijkl\}\{2211\}} \delta_{m1} 
-1 \delta_{\{ijkl\}\{2213\}} \delta_{m3} \right) \,, \nonumber \\
& & T^{30}_{ijklm} = \frac{1}{\sqrt{18}} \left( \delta_{\{ijkl\}\{2233\}} \delta_{m3} - \delta_{\{ijkl\}\{2231\}} \delta_{m1} \right) \,, \nonumber \\
& & T^{31}_{ijklm} = \frac{1}{\sqrt{72}} \left(3\delta_{\{ijkl\}\{2232\}} \delta_{m2} -2\delta_{\{ijkl\}\{2233\}} \delta_{m3} 
-1 \delta_{\{ijkl\}\{2231\}} \delta_{m1} \right) \,, \nonumber \\
& & T^{32}_{ijklm} = \frac{1}{\sqrt{18}} \left( \delta_{\{ijkl\}\{3311\}} \delta_{m1} - \delta_{\{ijkl\}\{3312\}} \delta_{m2} \right) \,, \nonumber \\
& & T^{33}_{ijklm} = \frac{1}{\sqrt{72}} \left(3\delta_{\{ijkl\}\{3313\}} \delta_{m3} -2\delta_{\{ijkl\}\{3311\}} \delta_{m1}   
-1 \delta_{\{ijkl\}\{3312\}} \delta_{m2} \right) \,, \nonumber \\
& & T^{34}_{ijklm} = \frac{1}{\sqrt{18}} \left( \delta_{\{ijkl\}\{3322\}} \delta_{m2} - \delta_{\{ijkl\}\{3321\}} \delta_{m1} \right) \,, \nonumber \\
& & T^{35}_{ijklm} = \frac{1}{\sqrt{72}} \left(3\delta_{\{ijkl\}\{3323\}} \delta_{m3} -2\delta_{\{ijkl\}\{3322\}} \delta_{m2}   
-1 \delta_{\{ijkl\}\{3321\}} \delta_{m1} \right) \,.
\end{eqnarray}
The 35-plet tensors obey the relation
\begin{eqnarray}
 \sum_{\mu = 1}^{35} T^{\mu}_{ijklm} T^{\mu}_{nopqr} & = 
& \frac{1}{24} \delta_{\{ijlk\}}\delta_{\{nopq\}}\delta_{mr} 
\nonumber \\
&& -\frac{1}{144} \left( \delta_{\{jkl\}}\delta_{\{opq\}}\delta_{im}\delta_{nr} + \delta_{\{jkl\}}\delta_{\{npq\}}\delta_{im}\delta_{or} \right.
\nonumber \\
&&       \qquad \quad + \delta_{\{jkl\}}\delta_{\{noq\}}\delta_{im}\delta_{pr} + \delta_{\{jkl\}}\delta_{\{nop\}}\delta_{im}\delta_{qr} 
\nonumber \\
&&       \qquad \quad + \delta_{\{ikl\}}\delta_{\{opq\}}\delta_{jm}\delta_{nr} + \delta_{\{ikl\}}\delta_{\{npq\}}\delta_{jm}\delta_{or} 
\nonumber \\
&&       \qquad \quad + \delta_{\{ikl\}}\delta_{\{noq\}}\delta_{jm}\delta_{pr} + \delta_{\{ikl\}}\delta_{\{nop\}}\delta_{jm}\delta_{qr} 
\nonumber \\
&&       \qquad \quad + \delta_{\{ijl\}}\delta_{\{opq\}}\delta_{km}\delta_{nr} + \delta_{\{ijl\}}\delta_{\{npq\}}\delta_{km}\delta_{or} 
\nonumber \\
&&       \qquad \quad + \delta_{\{ijl\}}\delta_{\{noq\}}\delta_{km}\delta_{pr} + \delta_{\{ijl\}}\delta_{\{nop\}}\delta_{km}\delta_{qr} 
\nonumber \\
&&       \qquad \quad + \delta_{\{ijk\}}\delta_{\{opq\}}\delta_{lm}\delta_{nr} + \delta_{\{ijk\}}\delta_{\{npq\}}\delta_{lm}\delta_{or} 
\nonumber \\
&&            \left. \qquad \quad + \delta_{\{ijk\}}\delta_{\{noq\}}\delta_{lm}\delta_{pr} + \delta_{\{ijk\}}\delta_{\{nop\}}\delta_{lm}\delta_{qr}  \right) \,.
\end{eqnarray}

\section{BUILDING THE OCTET OPERATORS}\label{app:operators}

The construction of operators relies on textbook group theory
methods (see, for example, Ref.~\cite{Cornwell}).
Table~\ref{table:JversusLambda} is a reminder of the connection between
angular momentum in the continuum and on a lattice\cite{Johnson,Berg:1982kp}.
To build an octet operator, begin with a single chair and
list all possible rotations of it.  There are 24 orientations in total,
as shown in Fig.~\ref{fig:chairs8}.
The $A_1$ representation is built from a particular sum,
\begin{equation}
H^{(8)\alpha}(A_1) = \left(\sum_{a=1}^{24}L^{(8)}_a\right)_{ij}T^\alpha_{ij}
\,.
\end{equation}
Any octahedral rotation of this sum leaves it invariant, as expected for a
$J=0$ operator.
The $A_2$ representation is built from a different sum,
\begin{equation}
H^{(8)\alpha}(A_2) = \left(
   \sum_{a=1}^{12}(-1)^aL^{(8)}_a - \sum_{a=13}^{24}(-1)^aL^{(8)}_a\right)_{ij}
   T^\alpha_{ij}
\,.
\end{equation}
Some octahedral rotations of this sum leave it invariant; others return
the negative of the sum.
The $T_1$ representation is built from a set of three sums,
\begin{eqnarray}
H^{(8)\alpha}(T_1^x)
 &=& \left(L^{(8)}_6 + L^{(8)}_{20} + L^{(8)}_{21} + L^{(8)}_{11}
         - L^{(8)}_{18} - L^{(8)}_8 - L^{(8)}_9 - L^{(8)}_{23}\right)_{ij}
     T^\alpha_{ij}\,, \\
H^{(8)\alpha}(T_1^y
) &=& \left(L^{(8)}_5 + L^{(8)}_{19} + L^{(8)}_{24} + L^{(8)}_{10}
          - L^{(8)}_{17} - L^{(8)}_7 - L^{(8)}_{12} - L^{(8)}_{22}\right)_{ij}
      T^\alpha_{ij}\,, \\
H^{(8)\alpha}(T_1^z)
 &=& \left(L^{(8)}_1 + L^{(8)}_2 + L^{(8)}_3 + L^{(8)}_4
         - L^{(8)}_{13} - L^{(8)}_{14} - L^{(8)}_{15} - L^{(8)}_{16}\right)_{ij}
     T^\alpha_{ij}\,.
\end{eqnarray}
Any octahedral rotation of one of these sums returns one of the three sums
(itself or one of the other two) up to $\pm1$, as expected for a vector
with $J=1$.
\begin{table}[hb]
\caption{The relationship between continuum angular momentum $J$
and octahedral irreducible representation $\Lambda$.}
\label{table:JversusLambda} 
\begin{tabular}{cccccc}
\hline
\hline
$\Lambda$ & \multicolumn{5}{c}{$J$} \\
\cline{2-6}
 & 0 & 1 & 2 & 3 & $\cdots$ \\
\hline
$A_1$ & 1 & 0 & 0 & 0 & $\cdots$ \\
$A_2$ & 0 & 0 & 0 & 1 & $\cdots$ \\
$E$   & 0 & 0 & 1 & 0 & $\cdots$ \\
$T_1$ & 0 & 1 & 0 & 1 & $\cdots$ \\
$T_2$ & 0 & 0 & 1 & 1 & $\cdots$ \\
\hline
\hline
\end{tabular}
\end{table}
The $T_2$ representation is built from a set of three sums,
\begin{eqnarray}
H^{(8)\alpha}(T_2^x)
 &=& \left(L^{(8)}_6 - L^{(8)}_{20} + L^{(8)}_{21} - L^{(8)}_{11}
         + L^{(8)}_{18} - L^{(8)}_8 + L^{(8)}_9 - L^{(8)}_{23}\right)_{ij}
     T^\alpha_{ij}\,, \\
H^{(8)\alpha}(T_2^y)
 &=& \left(L^{(8)}_5 - L^{(8)}_{19} + L^{(8)}_{24} - L^{(8)}_{10}
         + L^{(8)}_{17} - L^{(8)}_7 + L^{(8)}_{12} - L^{(8)}_{22}\right)_{ij}
     T^\alpha_{ij}\,, \\
H^{(8)\alpha}(T_2^z)
 &=& \left(L^{(8)}_1 - L^{(8)}_2 + L^{(8)}_3 - L^{(8)}_4
         + L^{(8)}_{13} - L^{(8)}_{14} + L^{(8)}_{15} - L^{(8)}_{16}\right)_{ij}
     T^\alpha_{ij}\,.
\end{eqnarray}
Any octahedral rotation of one of these sums returns one of the three sums
(itself or one of the other two) up to $\pm1$.
The $E$ representation is built from a set of three differences,
\begin{eqnarray}
v_1 - v_2\,, \\
v_2 - v_3\,, \\
v_3 - v_1\,,
\end{eqnarray}
where
\begin{eqnarray}
v_1 &\equiv& \left(
             L^{(8)}_6 + L^{(8)}_{20} + L^{(8)}_{21} + L^{(8)}_{11}
           + L^{(8)}_{18} + L^{(8)}_8 + L^{(8)}_9 + L^{(8)}_{23}
           \right)_{ij}T^\alpha_{ij}\,, \\
v_2 &\equiv& \left(
             L^{(8)}_5 + L^{(8)}_{19} + L^{(8)}_{24} + L^{(8)}_{10}
           + L^{(8)}_{17} + L^{(8)}_7 + L^{(8)}_{12} + L^{(8)}_{22}
           \right)_{ij}T^\alpha_{ij}\,, \\
v_3 &\equiv& \left(L^{(8)}_1 + L^{(8)}_2 + L^{(8)}_3 + L^{(8)}_4
           + L^{(8)}_{13} + L^{(8)}_{14} + L^{(8)}_{15} + L^{(8)}_{16}
           \right)_{ij}T^\alpha_{ij}\,.
\end{eqnarray}
Notice that only two of the three differences are linearly independent.
Any octahedral rotation of one of these differences returns one of the three
differences (itself or one of the other two) up to $\pm1$.

The next task is to calculate the character
$\chi$ of each representation, which is
defined to be the set of traces of the explicit matrix representation.
Since the octahedral group has five conjugacy classes, we need to evaluate
five traces per representation.  To be explicit, we can use
\begin{equation}
\left\{ \chi(e), \chi(c_2^{(xy)}), \chi(c_3^{(x\to y\to z\to x)}),
\chi(c_4^{(z)}), \chi\left(\left(c_4^{(z)}\right)^2\right) \right\}
\end{equation}
where $c_2^{(xy)}$ denotes a 180$^\circ$ rotation about the line $(x+y=1,z=0)$
and $c_3^{(x\to y\to z\to x)}$ denotes a 120$^\circ$ rotation about the line
$(x=y=z)$.

For the $A_1$, $A_2$, $T_1$, and $T_2$ representations, it is not necessary to
build an explicit matrix representation; we can merely sum the $\pm1$ factors
for those rotations that return an element to $\pm$(itself).
For the $E$ representation it is best to build the two-dimensional matrices
explicitly.  A convenient choice for the basis (as used, for example, in
Ref.~\cite{Petry:2008rt}) is
\begin{equation}
\left(\begin{array}{c} H^{(8)\alpha}(E^1) \\ H^{(8)\alpha}(E^2)
 \end{array}\right) =
\left(\begin{array}{c} \frac{1}{\sqrt{2}}(v_1 - v_2) \\
      \frac{-1}{\sqrt{6}}(v_1 + v_2 - 2v_3) \end{array}\right)
\end{equation}
and it leads to
\begin{eqnarray}
e &=& \left(\begin{array}{cc} 1 & 0 \\ 0 & 1 \end{array}\right)\,, \\
c_4^{(z)} &=& \left(\begin{array}{cc} -1 & 0 \\ 0 & 1 \end{array}\right)\,.
\end{eqnarray}
Under a $c_4^{(y)}$ rotation, we obtain
\begin{eqnarray}
\frac{1}{\sqrt{2}}(v_1-v_2) &\to& \frac{1}{\sqrt{2}}(v_3-v_2) \\
   &=& \frac{1}{2}\left(\frac{1}{\sqrt{2}}(v_1-v_2)\right)
    + \frac{\sqrt{3}}{2}\left(\frac{-1}{\sqrt{6}}(v_1 + v_2 - 2v_3)\right) \\
\frac{-1}{\sqrt{6}}(v_1 + v_2 - 2v_3)
     &\to& \frac{-1}{\sqrt{6}}(v_3 + v_2 - 2v_1) \\
   &=& \frac{\sqrt{3}}{2}\left(\frac{1}{\sqrt{2}}(v_1-v_2)\right)
    - \frac{1}{2}\left(\frac{-1}{\sqrt{6}}(v_1 + v_2 - 2v_3)\right)\,.
\end{eqnarray}
Therefore
\begin{equation}
c_4^{(y)} = \left(\begin{array}{cc} \frac{1}{2} & \frac{\sqrt{3}}{2} \\
            \frac{\sqrt{3}}{2} & -\frac{1}{2} \end{array}\right)\,.
\end{equation}
All other matrices can be obtained by multiplication of $c_4^{(y)}$ and
$c_4^{(z)}$ in various orders finding, for example,
\begin{eqnarray}
c_2^{(xy)} &=& \left(\begin{array}{cc} -1 & 0 \\ 0 & 1 \end{array}\right)\,,\\
c_3^{(x\to y\to z\to x)} &=&
      \left(\begin{array}{cc} -\frac{1}{2} & -\frac{\sqrt{3}}{2} \\
            \frac{\sqrt{3}}{2} & -\frac{1}{2} \end{array}\right) \,,\\
\left(c_4^{(z)}\right)^2
      &=& \left(\begin{array}{cc} 1 & 0 \\ 0 & 1 \end{array}\right) \,.
\end{eqnarray}
The characters of all representations are collected into
Table~\ref{table:characters},
\begin{table}[tb]
\caption{Characters $\chi$ of all irreducible representations $\Lambda$ for
the octet operator.}
\label{table:characters} 
\begin{tabular}{cccccc}
\hline
\hline
$\Lambda$ & $\chi(e)$ & $\chi(c_2^{(xy)})$ & $\chi(c_3^{(x\to y\to z\to x)})$ &
$\chi(c_4^{(z)})$ & $\chi((c_4^{(z)})^2)$ \\
\hline
$A_1$ & 1 & 1 & 1 & 1 & 1 \\
$A_2$ & 1 & --1 & 1 & --1 & 1 \\
$E$ & 2 & 0 & --1 & 0 & 2 \\
$T_1$ & 3 & --1 & 0 & 1 & --1 \\
$T_2$ & 3 & 1 & 0 & --1 & --1 \\
\hline
\hline
\end{tabular}
\end{table}
and the multiplicities are obtained through standard group theory methods:
\begin{eqnarray}
m(A_1) &=& \frac{1}{24}\left(\chi(e)+6\chi(c_2^{(xy)})
           +8\chi(c_3^{(x\to y\to z\to x)})+6\chi(c_4^{(z)})
           +3\chi((c_4^{(z)})^2)\right) = 1 \,, \\
m(A_2) &=& \frac{1}{24}\left(\chi(e)-6\chi(c_2^{(xy)})
           +8\chi(c_3^{(x\to y\to z\to x)})-6\chi(c_4^{(z)})
           +3\chi((c_4^{(z)})^2)\right) = 1 \,, \\
m(E) &=& \frac{1}{24}\left(2\chi(e)
           -8\chi(c_3^{(x\to y\to z\to x)})
           +6\chi((c_4^{(z)})^2)\right) = 1 \,, \\
m(T_1) &=& \frac{1}{24}\left(3\chi(e)-6\chi(c_2^{(xy)})
           +6\chi(c_4^{(z)})
           -3\chi((c_4^{(z)})^2)\right) = 1 \,, \\
m(T_2) &=& \frac{1}{24}\left(3\chi(e)+6\chi(c_2^{(xy)})
           -6\chi(c_4^{(z)})
           -3\chi((c_4^{(z)})^2)\right) = 1 \,.
\end{eqnarray}
To summarize, the operators that will be typed into the computer code are
those shown in Eq.~(\ref{eq:operators}).
Operators beyond the octet are built from this octet starting point, as
described in Sec.~\ref{sec:operators}.

\end{document}